\documentclass[twocolumn]{aastex631}
\usepackage {amsmath}
\usepackage {bm}
\usepackage {comment}
\usepackage{graphicx}

\newcommand{\Secref}[1]{Section~\ref{#1}}
\newcommand{\appref}[1]{Appendix~\ref{#1}}

\newcommand{\Eqref}[1]{Equation~(\ref{#1})}
\newcommand{\Figref}[1]{Figure~\ref{#1}}
\newcommand{\pr}[1]{\left({#1}\right)}

\shorttitle{GRB afterglow polarization}
\shortauthors{Kuwata et al.}

\graphicspath{{./}}

\begin{document}

\title{Synchrotron Polarization of Gamma-Ray Burst Afterglow Shocks with Hydrodynamic-scale Turbulent Magnetic Field}

\correspondingauthor{Asuka Kuwata}
\email{a.kuwata@astr.tohoku.ac.jp}

\author[0000-0002-6169-2720]{Asuka Kuwata}
\affiliation{Astronomical Institute, Graduate School of Science, Tohoku University, Sendai 980-8578, Japan}

\author[0000-0002-7114-6010]{Kenji Toma}
\affiliation{Frontier Research Institute for Interdisciplinary Sciences, Tohoku University, Sendai 980-8578, Japan}
\affiliation{Astronomical Institute, Graduate School of Science, Tohoku University, Sendai 980-8578, Japan}

\author[0000-0003-2579-7266]{Shigeo S. Kimura}
\affiliation{Frontier Research Institute for Interdisciplinary Sciences, Tohoku University, Sendai 980-8578, Japan}
\affiliation{Astronomical Institute, Graduate School of Science, Tohoku University, Sendai 980-8578, Japan}

\author[0000-0001-7952-2474]{Sara Tomita}
\affiliation{Frontier Research Institute for Interdisciplinary Sciences, Tohoku University, Sendai 980-8578, Japan}
\affiliation{Astronomical Institute, Graduate School of Science, Tohoku University, Sendai 980-8578, Japan}

\author[0000-0003-3383-2279]{Jiro Shimoda}
\affiliation{Department of Physics, Graduate School of Science, Nagoya University, Furo-cho, Chikusa-ku, Nagoya 464-8602, Japan}

\begin{abstract}
Afterglows of gamma-ray bursts (GRBs) are emitted from expanding forward shocks, which are expected to have magnetic field much stronger than the interstellar field, although the origin of the field is a long-standing problem. Two field amplification mechanisms, plasma kinetic instabilities and magnetohydrodynamic instabilities, have been discussed so far.
The coherence length scales of the fields amplified by these two processes are different by $7-10$ orders of magnitudes, and the polarimetric observations may distinguish them. We construct a semi-analytic model of the forward shock afterglow polarization
under the assumption of hydrodynamic-scale turbulent magnetic field.
We perform numerical calculations of synchrotron polarization for the isotropic turbulence and the zero viewing angle. We find that the polarization degrees are $\sim1~\%$ when the field coherence length scale in the fluid comoving frame is of the order of the thickness of the shocked regions. 
This range of polarization degree is comparable to that of the observed late-phase optical afterglows. Our model also shows that the radio polarization degrees are comparable to the optical ones on average but can be higher than the optical ones at some time intervals. The polarization angles are shown to vary randomly and continuously. These polarimetric properties are clearly different from the case of plasma kinetic  instability.
Simultaneous polarimetric observations of GRB afterglows at the radio and optical bands have recently started, which will help us constrain the magnetic field amplification mechanism.
\end{abstract}

\keywords{Gamma-ray bursts (629), Magnetic fields (994), Non-thermal radiation sources (1119), Jets (870)}

\section{Introduction} \label{sec:intro}

Gamma-ray bursts (GRBs) are intense flashes of gamma-rays from cosmological distances, and they are commonly understood as emission from relativistic jets launched after core collapse of massive stars or compact star mergers \citep{Hjorth2012, Abbott2017}. After the prompt gamma-ray emission, long-lasting broadband afterglows are usually observed. In the standard picture, the afterglows are non-thermal electron synchrotron emission from forward and reverse shocks formed by interaction of the relativistic jets with the ambient media \citep[][for reviews]{Meszaros2002, Piran2004, Kumar2015}. The reverse shock emission is bright in some bursts for $\sim 10^3\;$s at the optical band ($\sim 10^5\;$s at the radio band) \citep{Gao2015, Resmi2016}, and in other bursts or at later times
the forward shock emission is dominant, which we focus on in this paper.

At relativistic collisionless shocks, turbulent magnetic field is amplified and the wave-particle interaction gives rise to non-thermal electrons, but the detailed mechanisms of these processes are still elusive \citep[][for a review]{Sironi2013}. Understanding them will also help us extract more information on the properties of jets, progenitor systems, and their ambient media from observational data of afterglows. A field amplification mechanism widely discussed is Weibel instability at the shock, which amplifies magnetic field turbulence on the plasma skin depth scale \citep{Medvedev1999, 2005PhPl...12h0705K, Sironi2011, Ruyer2018, Takamoto2018, Lemoine2019},  although it is unclear whether this field can be maintained at a sufficiently long distance in the downstream that it can account for the observed synchrotron flux \citep{Gruzinov1999, Chang2008, Keshet2009, Tomita2016, Tomita2019, Asano2020, Huang2022}. Other instabilities that amplify turbulent field on hydrodynamic scales are also studied \citep{Sironi2007, InoueT2013, Mizuno2014, Duffell2014}.

Polarization of the synchrotron emission has a potential to be powerful probe for the magnetic field structure. For the late-phase forward shock afterglows, linear polarization degree (PD) has been measured for many bursts at the optical band at a level of PD $\sim 1-3 \%$ \citep{Covino2004, Covino2016}. 
The distribution of magnetic field can be anisotropic, depending on the angle from the shock normal, and then the net PD can be at the observed level even for the plasma-scale turbulent field model \citep{Sari1999, Ghisellini1999}. In this model, the optical polarization angle (PA) may flip by 90 degrees, although the PA flips were clearly observed only in GRB 121024A and GRB 091018 \citep{Wiersema2014}.\footnote{Jets with angular structure and forward shocks with strong ordered magnetic field can have no PA flips \citep{Rossi2004, Granot2003}. The optical observation of GRB 020813 shows no PA flip \citep{Lazzati2004}.} This model also predicts low PD in the phase earlier than the jet break time (say, $\sim 1$ day). It is disfavored by the detection of early-phase forward shock polarization, PD $\sim 10 \%$, in GRB 091208B with Kanata Telescope \citep{Uehara2012}, although there might be contribution from highly-polarized reverse shock emission \citep{Jordana-Mitjans2021, Mundell2013}.

Recently, radio forward shock polarization in the optically thin regime for synchrotron self-absorption has been first detected with ALMA in GRB 171205A \citep{Urata2019}. Simultaneous observations at the optical and radio bands will be a new observational test of the magnetic field structure. The radio PD is lower than the optical one generally in the plasma-scale turbulent field model \citep{ST2021, Birenbaum2021, Rossi2004}.

In contrast to the plasma-scale turbulent field model, the hydrodynamic-scale turbulent field model has not been studied well. It has been expected that the PD and PA temporally change in a random manner and PD $\sim 70/\sqrt{N} \%$, where $N$ is the number of patches within which the magnetic field is considered to be coherent in the relativistically-beamed visible region \citep{Gruzinov1999}, but no numerical calculations have been shown so far. Frequency dependence of PD and PA has not been studied, either. In this paper, we build a semi-analytic model of hydrodynamic-scale turbulent field to predict its multi-band polarimetric signature. We show that the radio PD can be higher than the optical one, differently from the plasma-scale field model. 

This paper is organized as follows. In \Secref{sec:blast}, we introduce the dynamics and synchrotron emission from the expanding forward shocks. 
In \Secref{sec:large-B}, we construct a synchrotron polarization model with hydrodynamic-scale turbulent magnetic field.
We analytically estimate the level of the polarization degree in \Secref{sec:analytical-estimate} and we show the numerical calculation results of multi-wave band synchrotron polarization in \Secref{sec:results}.
\Secref{sec:summary} is devoted to summary and discussion.

\section{Forward shock dynamics and emission flux} \label{sec:blast}
We consider GRB afterglows as synchrotron emission from relativistically and adiabatically expanding spherical forward shocks. We calculate their dynamics and emission fluxes by following the formulation of \cite{Granot1999a} with the thin-shell approximation (see Section \ref{subsec:flux}). We also take into account the collimation of outflows. Throughout this paper, the superscript prime denotes a physical value measured in the comoving frame (i.e., the rest frame of the fluid in the downstream of the shock wave).

\subsection{Equal arrival time surface} \label{subsec:EATS}
The radius of the shock front is denoted by $R=R(t)$, and its Lorentz factor $\Gamma$ scales as $R^{-3/2}$ in the case of adiabatic expansion \citep{BM1976}. The radius $R$ can be rewritten as a function of a photon arrival time to the observer, $T$. We use a spherical coordinate system centered on the shock wave and set our line of sight along the $z$-axis for convenience. For a photon emitted at time $t$ and position $(r,\mu)$ in the lab frame, where $\mu \equiv \cos \theta$, the arrival time $T$ is 
\begin{equation}
    \frac{T}{1+z} = t - \frac{r\mu}{c},
    \label{eq:observer-time}
\end{equation}
where $z$ is the cosmological redshift and $c$ is the speed of light. 
We have chosen $T=0$ as the arrival time of a photon emitted from the origin at time $t=0$. Solving the motion equation of the shock, $dR/cdt = \sqrt{1-1/\Gamma^2}$ with $\Gamma \propto R^{-3/2}$ and $\Gamma \gg 1$, we obtain
 \begin{equation}
    \label{eq:R-EATS}
     R \simeq \frac{c T/(1+z)}{1 - \mu + 1/(8\Gamma^2)}.
 \end{equation}
This surface is called ``the equal arrival time surface (EATS)''.

\begin{figure}[t]
    \centering
    \includegraphics[width=8cm]{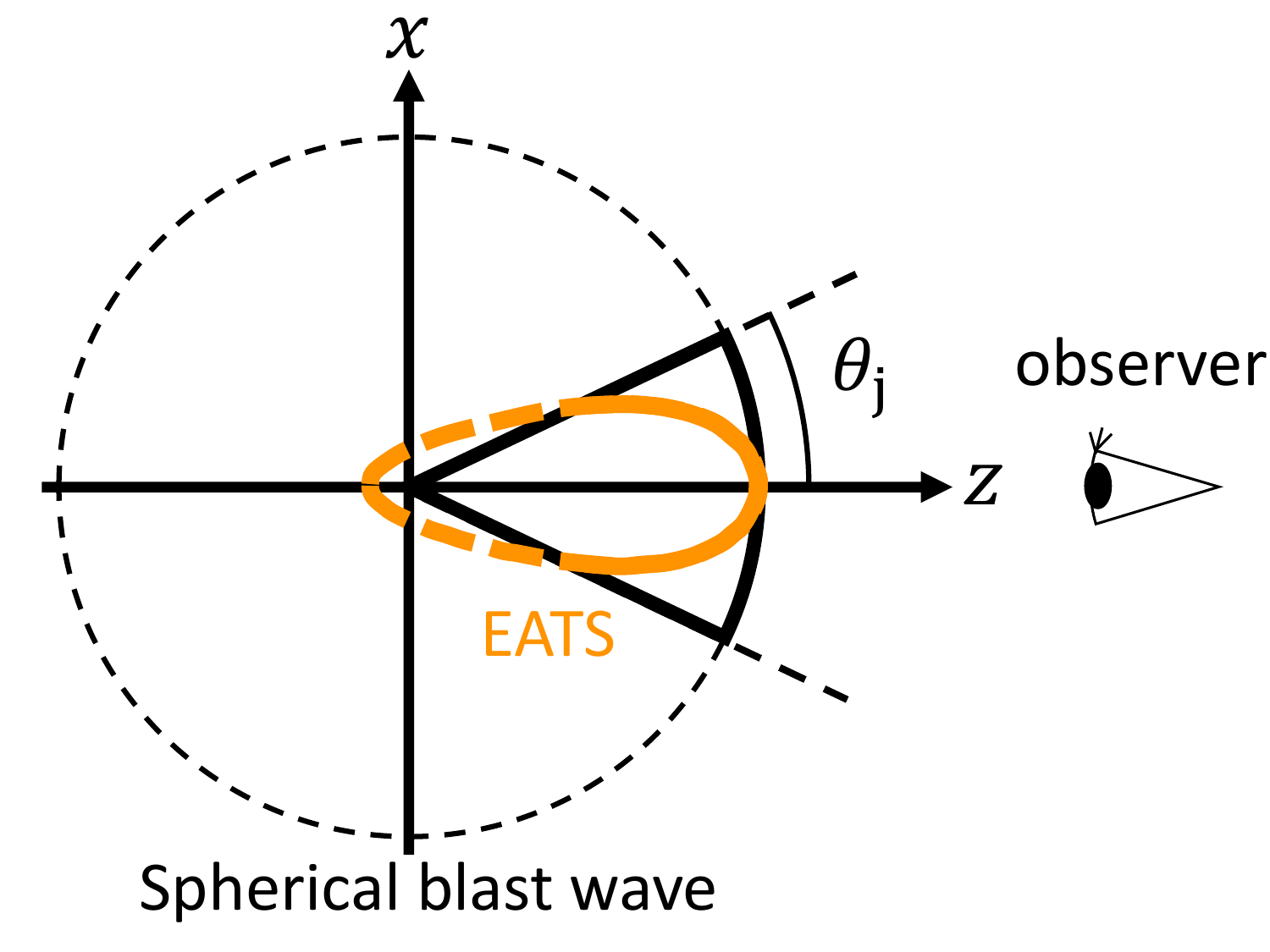}
    \caption{Schematic diagram of a forward shock. The black dashed circle corresponds to the expanding spherical shock wave. A GRB jet produces the collimated shock wave which has the opening angle $2\theta_{\rm j}$ (black thick solid sector). A line of sight is fixed along the $z$-axis. The orange egg-like shape (EATS) is the emission region which has the same arrival time $T$ in the thin-shell approximation. No emission comes from the orange dashed line due to the collimation of outflow. \label{fig:geometry}}
\end{figure}

\Figref{fig:geometry} schematically shows the geometry of our system. We regard a part of the blast wave within the angle interval $2\theta_j$ as produced by a GRB jet, and exclude the other part. In this work, we focus on the case of the viewing angle $\theta_v=0$. 

\subsection{Emission flux} \label{subsec:flux}
The energy flux density of synchrotron emission should be considered based on EATS. \cite{Granot1999a} provided with a general formula of the flux density of radiation from a spherical expanding system in the case of the optically-thin limit and isotropic radiation in the comoving frame,
\begin{align}
    F(\nu,T) = \frac{1+z}{4\pi d_L^2} \int_0^{2\pi} d\phi \int_{-1}^1 d\mu \int_0^\infty r^2 dr
               \frac{P'(\nu', \bm{r}, t)}{\gamma^2(1-\beta \mu)^2},
    \label{eq:flux-general}
\end{align}
where $d_L$ is the luminosity distance to a GRB, $\gamma$ is the Lorentz factor of the fluid, $\beta = \sqrt{1-1/\gamma^2}$, $\nu' = \gamma \nu (1-\beta \mu)$, and $t$ is given by \Eqref{eq:observer-time}. The factor $1/\gamma^2(1-\beta\mu)^2$ represents the Doppler beaming effect, by which the bright region is concentrated to $\theta \lesssim \gamma^{-1}$. The emission from the region at $\theta > \gamma^{-1}$ is beamed away from the line of sight. We set the emission power $P'$ to be zero to $\cos \theta < \cos \theta_j$ (see \Figref{fig:geometry}). 

$P'$ depends on the number and energy densities in the shocked region, which rapidly decrease on the length scale of $\sim R/\Gamma^2$ in the downstream of the shock front. This structure of the shocked region with $\Gamma \gg 1$ is given by Eqs. (27)-(30) and (40)-(42) of \cite{BM1976}, and we call it ``BM structure''. In our model, instead of taking account of BM structure, we approximate the bright region behind the shock front as a thin shell. The number density and internal energy density of the thin shell are then given by
\begin{align}
        n' &= 4\gamma_f n, \label{eq:thinshell-n} \\
        e' &= 4\gamma_f^2 n m_{\rm p} c^2, \label{eq:thinshell-e} 
\end{align}
where $\gamma_f$ is the Lorentz factor of the fluid just behind the shock front, $n$ is the number density of ambient medium, and $m_{\rm p}$ is the proton mass.

We employ the thin-shell approximation as
\begin{align}
    P'(\nu',\bm{r},t) = P'(\nu',R,t) &\times \Delta R' \times C \nonumber\\
                                     &\times \delta (r'-R'(t')),
    \label{eq:power-thinshell}
\end{align}
where $\Delta R' = R(t)/16\gamma_f$ is the thickness of shocked region in the comoving frame, and $C$ is the normalization constant to fit the approximated flux to the flux calculated with BM structure.
We introduce a self-similar variable $y$ 
\begin{equation}
    y \equiv \frac{R}{R_l},
\end{equation}
where $R_l$ is the radius of the shock from which a photon on the line of sight reaches the detector at a time $T$.
From \Eqref{eq:R-EATS}, $y$ depends on $T,r$ and $\mu$. 
We can also describe the condition of adiabatic expansion as
\begin{equation}
    \gamma_f = \gamma_l y^{-3/2},
\end{equation}
where $\gamma_l$ is the Lorentz factor of the fluid just behind the shock at $R = R_l$.
From \Eqref{eq:observer-time} and $R_l = 16\gamma_l^2 c T/(1+z)$, we can express $r, \mu$ in terms of $y$ \citep{Granot1999a},
\begin{equation}
    \label{eq:r-mu}
    r = R = R_l y,\quad \mu \simeq 1-\frac{1 - y^4}{16\gamma_l^2 y}.
\end{equation}
Substituting \Eqref{eq:power-thinshell} and \Eqref{eq:r-mu} into \Eqref{eq:flux-general}, we obtain     
\begin{equation}
    F(\nu,T) = C \frac{64(1+z)R_l^3}{\pi d_L^2} \int_0^{2\pi} d\phi \int_{\frac{1}{32\gamma_l^2}}^1 dy \frac{(1+3y^4) y^{10} P'}{(1+7y^4)^3}.
    \label{eq:flux-thinshell}
\end{equation}
See \Secref{subsec:param} for the value of $C$.

\subsection{Synchrotron power} \label{subsec:syn}
To calculate $P'$ at each point, we assume that the energy densities of accelerated electrons $e'_e$ and magnetic field $e'_B$ are fixed fractions of the local internal energy; $e'_e = \epsilon_e e'$ and $e'_B = \epsilon_B e'$.
We also assume that the electrons are accelerated by the forward shock and they have a single power-law distribution function everywhere in the downstream of the shock:
\begin{equation}
    N(\gamma'_e) = K {\gamma'_e}^{-p}\quad {\rm for}\ \gamma'_e \ge \gamma'_m,
\end{equation}
where $\gamma'_e$ is electron's Lorentz factor and $p$ is constant ($p>2$).
The normalization constant $K$ is determined as 
\begin{equation}
    K = (p-1) n' {\gamma'_m}^{p-1},
\end{equation}
and the minimum Lorentz factor of electrons $\gamma'_m$ is determined as
\begin{equation}
    \gamma'_m = \pr{\frac{p-2}{p-1}} \frac{\epsilon_e e'}{n' m_e c^2},
\end{equation}
where $m_e$ is electron mass. 

\begin{deluxetable}{lrl}
	\tablecaption{Parameters in this work \label{table:ordered-param}}
	\centering
	\tablehead{
    \colhead{parameter} & \colhead{symbol} & \colhead{value}
    }
    \startdata
		redshift & $z$ & $0.0$ \\
	    isotropic energy of blast wave & $E_{\rm iso}$ & $2.0\times 10^{52}\ {\rm erg}$  \\
		upstream number density & $n$ & $1.0\ {\rm cm^{-3}}$ \\
		power-law index of accelerated electrons & $p$ & $2.5$ \\
		energy fraction of accelerated electrons & $\epsilon_e$ & $0.1$ \\
		energy fraction of magnetic field & $\epsilon_B$ & $0.01$ \\
		opening angle of jet & $\theta_j$ & $6.0\ {\rm deg.}$ \\
	    viewing angle of jet & $\theta_v$ & $0.0\ {\rm deg.}$ \\
	\enddata
\end{deluxetable}

\cite{Granot1999a} gives approximate synchrotron power formulas, 
\begin{equation}
    P' =
    \begin{cases}
        P'_{\rm \nu',max} \pr{\frac{\nu'}{\nu'_m}}^{\frac{1}{3}} & {\rm for}\ \nu' < \nu'_m \\
        P'_{\rm \nu',max} \pr{\frac{\nu'}{\nu'_m}}^{-\frac{p-1}{2}} & {\rm for}\ \nu' > \nu'_m
    \end{cases},
    \label{eq:synpower-approx}
\end{equation}
where
\begin{equation}
    P'_{\rm \nu',max} = 0.88 \frac{4(p-1)}{3p-1} \frac{n' P'_{\rm e,avg}}{\nu'_{\rm syn}(\langle \gamma'_e \rangle)}
\end{equation}
where a factor of $0.88$ comes from the fitting with exact synchrotron emission and $P'_{\rm e,avg}$ is the synchrotron power by a single electron with an average Lorentz factor of $\langle \gamma'_e \rangle \equiv \epsilon_e e'/(n' m_e c^2)$. This is given as 
\begin{equation}
    P'_{\rm e,avg} = \frac{4}{3} \sigma_T c \beta_e^2 \langle \gamma'_e \rangle^2 \epsilon_B e',
\end{equation}
where $\sigma_T$ is the Thomson cross section and $\beta_e = \sqrt{1-1/\langle \gamma'_e \rangle^2}$.
The synchrotron peak frequency in the fluid frame is 
\begin{equation}
    \nu'_{\rm syn} (\gamma'_e) = \frac{3\gamma_e^{'2} q_e B'}{16 m_e c},
\end{equation}
where $q_e$ is an electron's electric charge and $B' = \sqrt{8\pi \epsilon_B e'}$ is the local magnetic field strength. $\nu'_m$ in \Eqref{eq:synpower-approx} is defined as $\nu'_m \equiv \nu'_{\rm syn} (\gamma'_m)$.

\subsection{Parameter setting, light curves and brightness profile} \label{subsec:param}
\begin{figure}[t]
    \centering
    \includegraphics[width=8.6cm]{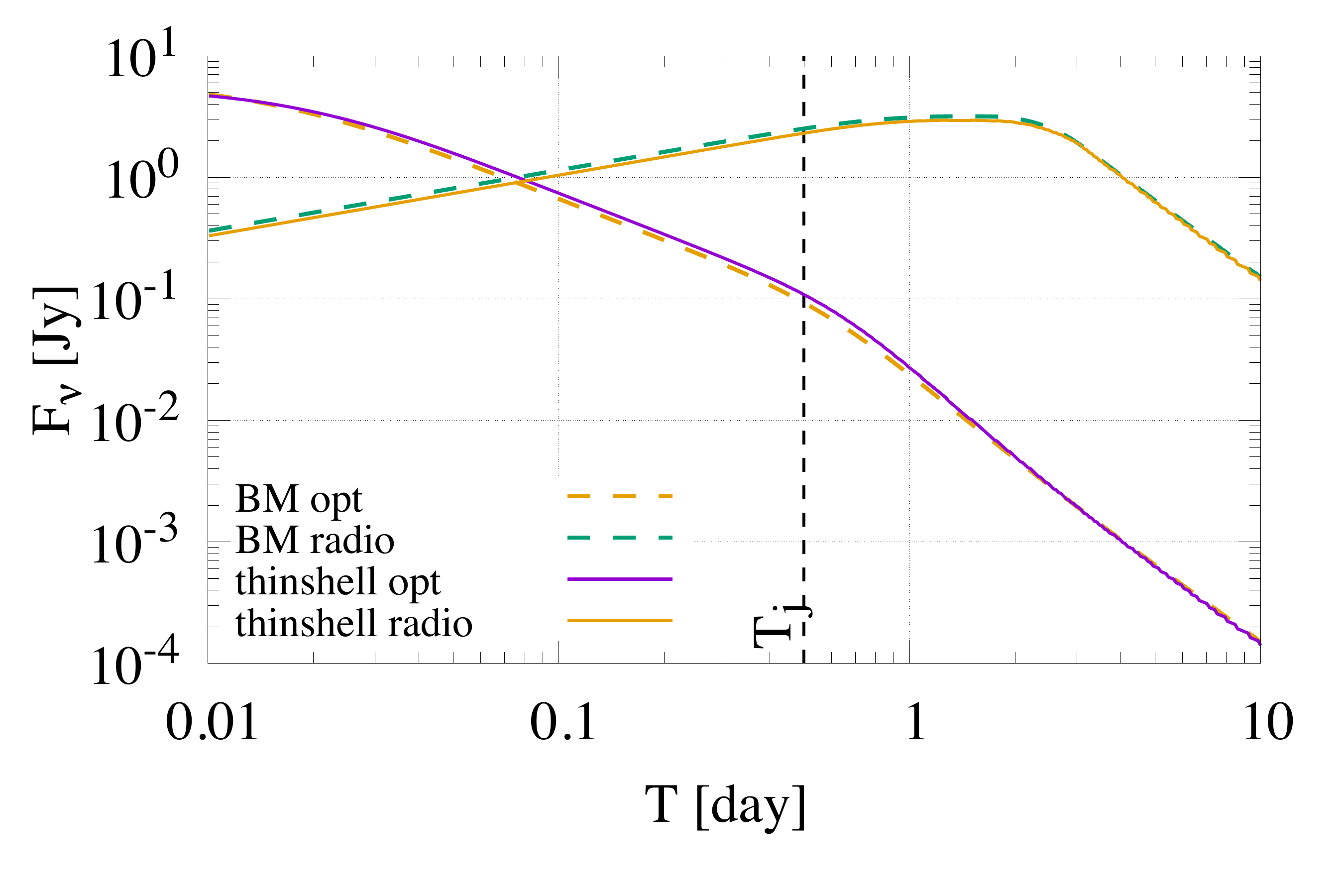}
    \caption{Synchrotron light curves at frequencies $10^{15}\;$Hz (optical) and $100\;$GHz (radio) with the thin-shell approximation (solid lines) and with the BM structure (dashed lines). The dashed vertical lines indicate the jet break time $T_j$.}
    \label{fig:light-curve}
\end{figure}

The Lorentz factor of the shock wave is given by 
\begin{equation}
    \Gamma = \frac{1}{2} \pr{\frac{17E_{\rm iso}}{16\pi n m_{\rm p} c^5 T^3/(1+z)^3}}^{1/8},
\end{equation}
according to \cite{BM1976}. Then, we can obtain observed synchrotron flux by using the above formulae and plot the light curves in \Figref{fig:light-curve}. For this calculation we have used the parameter values shown in Table 1, which are typical ones of long GRBs \citep[e.g.][]{Panaitescu2002}, except for the redshift. We assume nearby events with $z \sim 0$. We fix those parameter values for the flux calculation and change a parameter value on the turbulent magnetic field to study the properties of PD and PA (see \Secref{sec:large-B}). 

We consider that the sideways expansion of collimated relativistic blast waves is weak, as indicated by high-resolution hydrodynamic simulations \citep{Zhang2009, vanEerten2012}, and thus we set $\theta_j =$ const. for simplicity. As we mentioned in \Secref{subsec:EATS}, we focus on the case of $\theta_v = 0$. In \Figref{fig:light-curve}, the light-curve break at $T\simeq0.5~{\rm day}$ is due to the collimation of outflow (i.e., jet break).

\begin{figure*}[htbp]
    \centering
    \includegraphics[width=14cm]{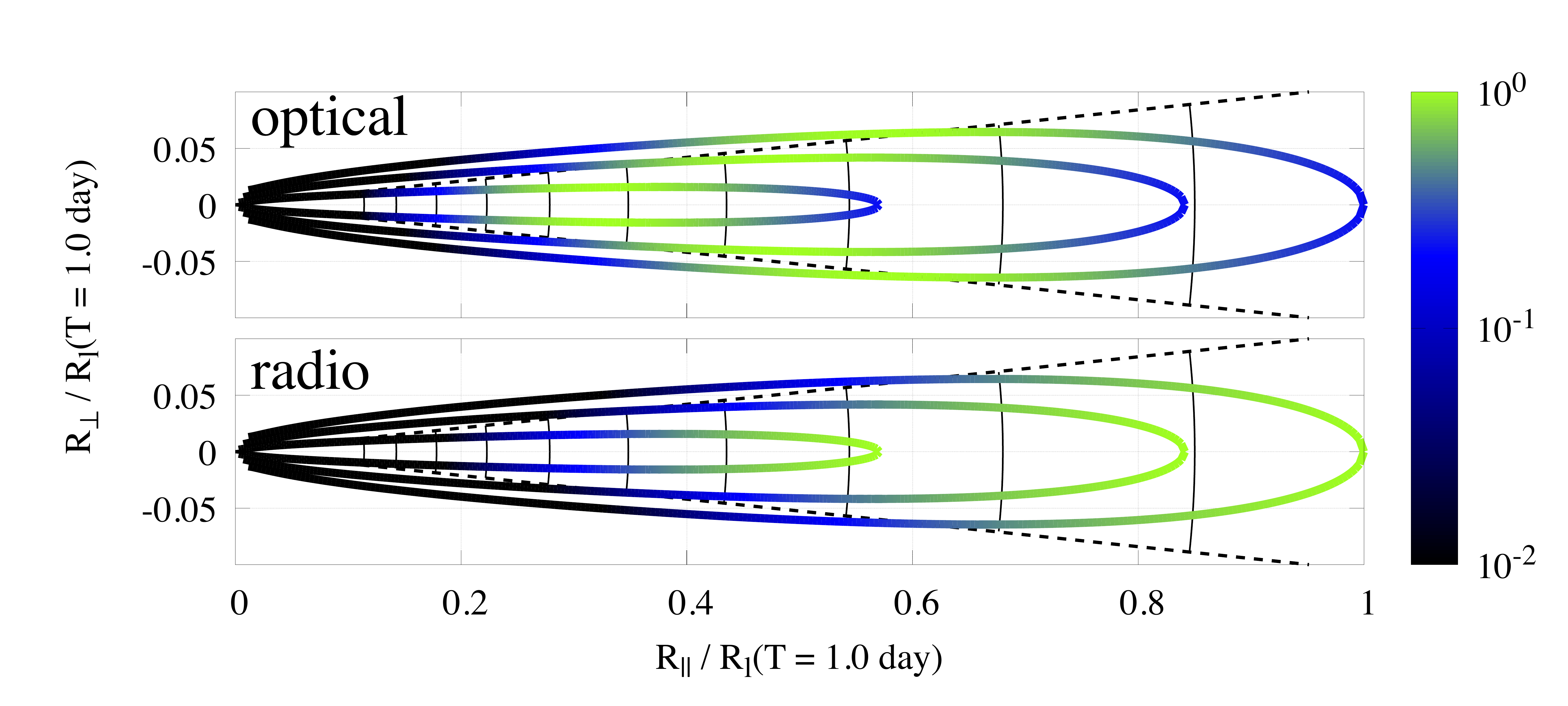}
    \caption{The flux density element $dF/dyd\phi$ (in arbitrary units) at the frequency $10^{15}~{\rm Hz}$ (optical, the upper panel) and $100~{\rm GHz}$ (radio, the lower panel) on EATS at $T=0.1~{\rm days}$ (the smallest one), at $T=0.5~{\rm days}$ (the middle one) and at $T=1.0~{\rm days}$ (the largest one). The dashed black lines indicate $\theta = \theta_{\rm j}$. 
    The black solid lines describe spheres separated by the spatial interval $\delta R_{B}$ in \Eqref{eq:delta-R-b}, which represent the thin shells with different configurations of turbulent magnetic field. See the texts in \Secref{subsec:B-field} for details.
    \label{fig:flux-density-element}}
\end{figure*}

According to \Eqref{eq:synpower-approx}, the integrated flux spectrum has a broken power-law form, and its peak frequency is given by 
\begin{equation}
    \nu_T = \nu_m (y=1).
\end{equation}
We consider the frequencies of $10^{15}~{\rm Hz}$ (optical) and $100~{\rm GHz}$ (radio). In the time interval $T = 0.1-1\;$day, for which we calculate PD and PA, $\nu_T$ always lies between the radio and optical frequencies. 
We have confirmed that the synchrotron cooling and synchrotron self-absorption effects are not important for our parameter set, similarly to the case in \cite{ST2021}, i.e., $\nu_a < 100~{\rm GHz} < \nu_T < 10^{15}~{\rm Hz} < \nu_c$. The break of the radio light curve at $T\simeq 2\;$day is due to the crossing of $\nu_T$.

The dashed lines in \Figref{fig:light-curve} represent the calculation results with BM structure. We find that the thin-shell approximation can mimic the emission flux from the shock with BM structure very well with $C = 0.80$. 

In \Figref{fig:flux-density-element}, we show the flux density element $dF/dyd\phi$, which is the integrand of \Eqref{eq:flux-thinshell} at the optical and radio bands. At the optical band, the brightest part is around the middle part of the EATS, $y \sim 0.5-0.7$, while at the radio band, it is on the observer side of the EATS, $y \sim 0.8-1.0$. This frequency dependence of the brightness profile causes the difference of the polarization between the optical and radio bands, as shown in \Secref{sec:analytical-estimate} and \ref{sec:results}.

\section{polarization from hydrodynamic-scale turbulent magnetic field} \label{sec:large-B}
Synchrotron polarization depends on the magnetic field configuration at each local position. We consider the turbulent magnetic field with coherence length of hydrodynamic scale, which may be produced by magnetohydrodynamic instabilities such as Richtmyer-Meshkov instability \citep{Sironi2007,InoueT2013, Mizuno2014}. In this case, the maximum coherence length of the magnetic field will be $\sim \Delta R$. Since the detailed structure of magnetic field in the shocked region has not been clear yet, we build a generic, semi-analytic model with parametrized field anisotropy under the assumption that the wavelength of turbulent field is $\sim \Delta R$.
In reality, the turbulent magnetic field will be amplified by the instabilities on a few eddy turnover times and cascade to smaller scales in the shocked region, forming a power spectrum of the magnetic energy \citep{Brandenburg2005, InoueT2011, Xu&Lazarian2016, Tomita2022}. 
We confirm that our conclusion on the synchrotron polarization does not change even in a case of Kolmogorov-type turbulence (see \appref{app:Kolmogorov-spectrum}). The distribution of magnetic field strength and coherence length in the shocked region will depend on the amplification time scale, but we ignore the amplification process for simplicity.

\subsection{Turbulent magnetic field model} \label{subsec:B-field}
We use a method based on the work of \cite{Giacalone1999} and its application to the polarization of supernova remnants \citep{Bykov2020}. To obtain the turbulent magnetic field that varies on the thin shell, we first derive turbulent magnetic field on a 2D plane and then rotate it onto the spherical thin shell. 

We consider the summation of a large number of waves with 
directions of wavevectors and phases given by the uniform random numbers
on a 2D plane whose normal is along $\bm{\hat{z}}$:
\begin{equation}
    \bm{B}(x,y) = \sum_{\rm n=1}^{N_m} \bm{\hat{b}}_n \exp{(i\bm{k}_n \cdot \bm{r}'_n + i\beta_n)},
    \label{eq:B-field}
\end{equation}
where
\begin{equation}
    \label{eq:B-field-direction}
    \bm{\hat{b}}_n = \sigma_{\perp} \cos \alpha_n \bm{\hat{y}}'_n + i\ \sigma_{\rm \|} \sin \alpha_n \bm{\hat{z}},
\end{equation}
the transformation from $\bm{r} = (x,y,z)$ to $\bm{r}'_n = (x'_n,y'_n,z)$ is described by
\begin{equation}
	\begin{pmatrix}
		x^\prime_n\\
		y^\prime_n
	\end{pmatrix}
	=
	\begin{pmatrix}
		\cos \phi_n & \sin \phi_n  \\
		-\sin \phi_n & \cos \phi_n  \\
	\end{pmatrix}
	\begin{pmatrix}
		x\\
		y
	\end{pmatrix},
\end{equation}
and $\phi_n, \alpha_n, \beta_n$ are random numbers. We have two orthogonal directions $\bm{\hat{y}}'_n, \bm{\hat{z}}$ of the turbulent magnetic field, which are chosen orthogonal to the wavevector $\bm{k}_n \| \bm{\hat{x}}'_n$ in order to satisfy $\bm{\nabla} \cdot \bm{B} = 0$ in the lab frame. $\sigma^2_{\perp}$ and $\sigma^2_{\parallel}$ are the variances of the wave amplitude in $\bm{\hat{y}}'_n$ and $\bm{\hat{z}}$ directions, respectively. We assume that the turbulent magnetic field is isotropic in the comoving frame,
    \begin{equation*}
        \frac{2{\sigma'_\|}^2}{{\sigma'_\perp}^2} = 1,
    \end{equation*}
so that we have $\sigma_\perp^2 = 2 \Gamma^2 \sigma_\|^2$. For simplicity, we set $|\bm{k}_n| = k_0 =$ const. Let $\lambda'_B$ denote the wavelength corresponding to $k_0$ in the comoving frame. We assume that $\lambda'_B$ is of the order of $\Delta R' \simeq R/16\gamma_f$, and write
\begin{equation}
    \lambda'_{B} = f_B \frac{R}{16\gamma_f}, 
    \label{eq:lambda-B}
\end{equation}
where $f_B$ is a parameter. 
(We also calculate polarization for Kolmogorov power spectrum in \appref{app:Kolmogorov-spectrum}).
    
Finally we model the radial variation of the magnetic field on the expanding thin shell to calculate the polarization of emission from EATS. Consider the radial length scale $\lambda_B^{\rm sh}$, which is the field wavelength measured in the rest frame of the shock. Then the timescale for which the shocked fluid advect from the shock front to the length scale of $\lambda_B^{\rm sh}$ is
\begin{equation}
    \delta t^{\rm sh}_B \simeq \frac{\lambda_B^{\rm sh}}{c/3} = 2\sqrt{2} \frac{\lambda'_B}{c}.
    \label{eq:delta-t-prime}
\end{equation}
In the same timescale, new turbulent eddies amplifying the field of scale $\sim \lambda_B^{\rm sh}$ will be created behind the shock front. Although the turbulent field changes moment by moment in reality, we assume that the turbulent field under our thin-shell approximation changes \textit{in a discrete manner} for simplicity: The field structure of the thin shell is set to be unchanged during the timescale $\delta t_B^{\rm sh}$ and reset with new realization of the random numbers at every $\delta t_B^{\rm sh}$. 
The timescale $\delta t_B^{\rm sh}$ is measured in the lab frame as $\delta t_B = \Gamma \delta t^{\rm sh}_B \simeq \sqrt{2} \gamma_f \delta t^{\rm sh}_B$, for which the thin shell expands over the distance of
\begin{equation}
    \delta R_B = c \delta t_B \simeq \frac{1}{4} f_B R.
    \label{eq:delta-R-b}
\end{equation}
The black solid lines in \Figref{fig:flux-density-element} represent the spheres with the spatial interval $\delta R_B$, which are set to have different configurations of turbulent magnetic field. The field configurations are unchanged between these spheres.

\subsection{Synchrotron polarization} \label{subsec:polari}
To calculate the polarization of synchrotron emission, we first derive the electric field $\bm{e}'$ of radiation in the comoving frame by $\bm{e}' = \bm{B}' \times \bm{n}'$ \citep[cf.][]{2003ApJ...597..998L,2009ApJ...698.1042T}. $\bm{n}'$ is the unit wavevector of observed radiation in the comoving frame, which is related to $\bm{\hat{n}} \parallel \bm{\hat{z}}$ as
         \begin{equation}
             \bm{\hat{n}}' = \frac{ \bm{\hat{n}} + \Gamma \bm{\beta}
 			(\frac{\Gamma}{\Gamma + 1}\bm{\hat{n}} \cdot \bm{\beta} - 1) }{ \Gamma (1 - \bm{\hat{n}} \cdot \bm{\beta} ) },  
 			\label{eq:n-prime}
 		\end{equation}
where $\bm{\beta} = \beta \bm{\hat{r}}$ (i.e., the spherical expansion). The magnetic field in the comoving frame is given by $\bm{B}_\parallel' = \bm{B}_\parallel$ and $\bm{B}_\perp' = \bm{B}_\perp/\Gamma$ under the ideal MHD approximation. Then we obtain the radiation electric field in the lab frame by Lorentz transformation,
     \begin{equation}
         \bm{e} = \Gamma \left[  \bm{e}' - \frac{\Gamma}{\Gamma+1}
 	             (\bm{e}' \cdot \bm{\beta}) \bm{\beta} - \bm{\beta}\times
 	             (\bm{n}' \times \bm{e}') \right].
     \end{equation}

The PA $\phi_p$ at each grid of EATS is given by
\begin{equation}
    \phi_p = \arctan \pr{\frac{e_y}{e_x}}.
\end{equation}
We obtain observed Stokes parameters $I_\nu, Q_\nu, U_\nu$, which are integrated over EATS:
\begin{align}
	&I_\nu = \int \frac{dF}{dyd\phi} dyd\phi, \label{eq:I-nu} \\
	&Q_\nu = \int \Pi_0 \cos(2\phi_p) \frac{dF}{dyd\phi} dyd\phi, \label{eq:Q-nu} \\
	&U_\nu = \int \Pi_0 \sin(2\phi_p) \frac{dF}{dyd\phi} dyd\phi, \label{eq:U-nu}
\end{align}
where $\Pi_0$ is the synchrotron PD in the  uniform magnetic field, which has frequency dependence \citep{Radipro,Melrose1980b},
\begin{equation}
	\label{eq:Pi-0}
	\Pi_0 = 
	\begin{cases}
		0.5 & (\nu < \nu_m) \\
		\frac{p+1}{p+7/3} & (\nu \geq \nu_m) 
	\end{cases}.
\end{equation}
Note that we focus on the spectral segments at $\nu_a < \nu < \nu_c$ (see Section \ref{subsec:param}). 
We obtain the net PD $\Pi_\nu$ and the net PA $\Phi_{p,\nu}$ using $I_\nu, Q_\nu, U_\nu$:
\begin{align}
	&\Pi_\nu = \frac{\sqrt{Q_\nu^2 + U_\nu^2}}{I_\nu}, \label{eq:pi-tot} \\
	&\Phi_{p,\nu} = \frac{1}{2} \tan^{-1}\pr{\frac{U_\nu}{Q_\nu}}. \label{eq:phi-tot}
\end{align}
We write the net PDs at $10^{15}$ Hz and $100$ GHz by $\Pi_{\rm opt}$ and $\Pi_{\rm radio}$, respectively.

\section{Analytical estimate of PD} \label{sec:analytical-estimate}

\begin{figure*}[t]
    \centering
    \includegraphics[width=8.5cm]{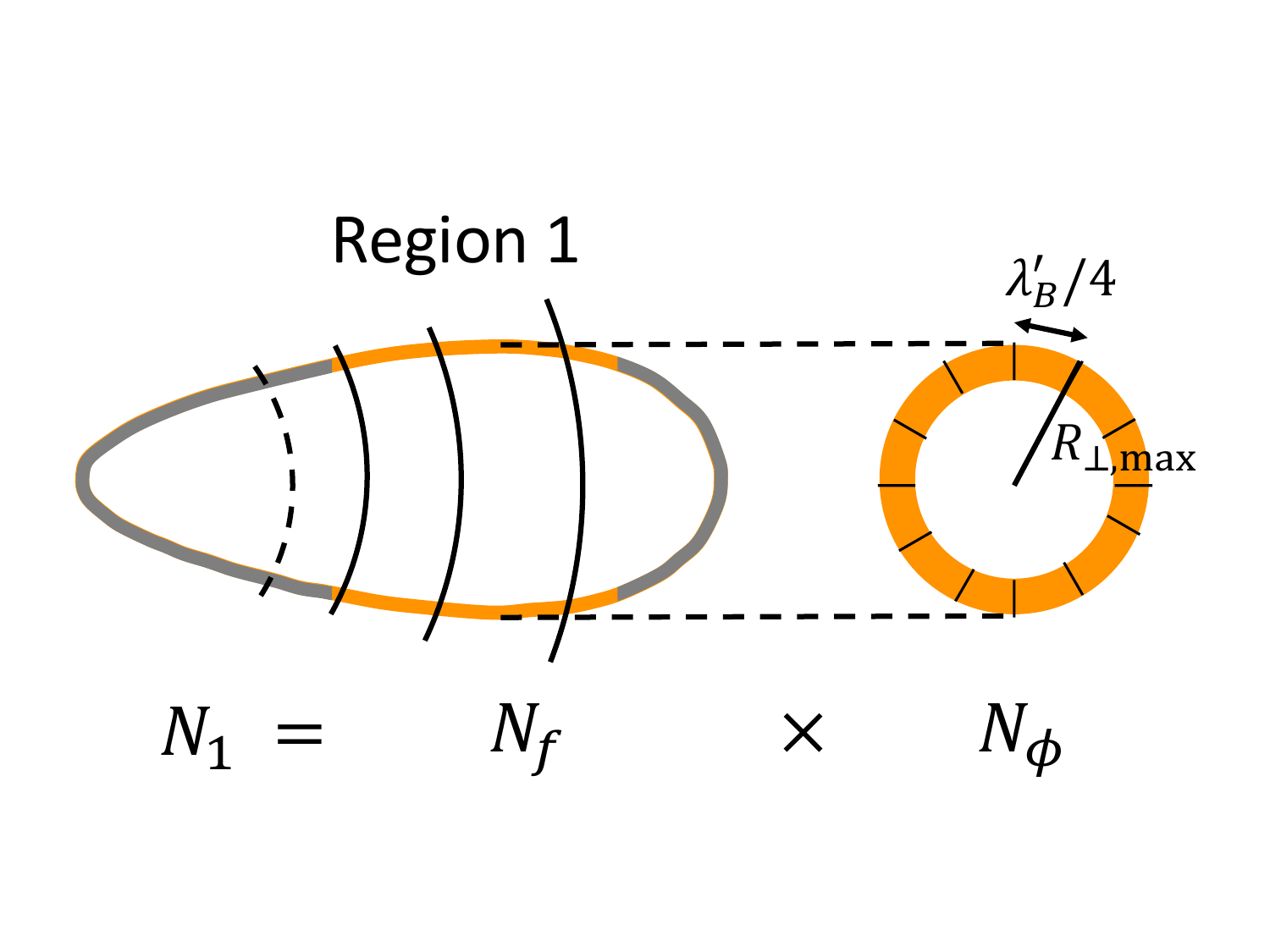}
    \includegraphics[width=8.5cm]{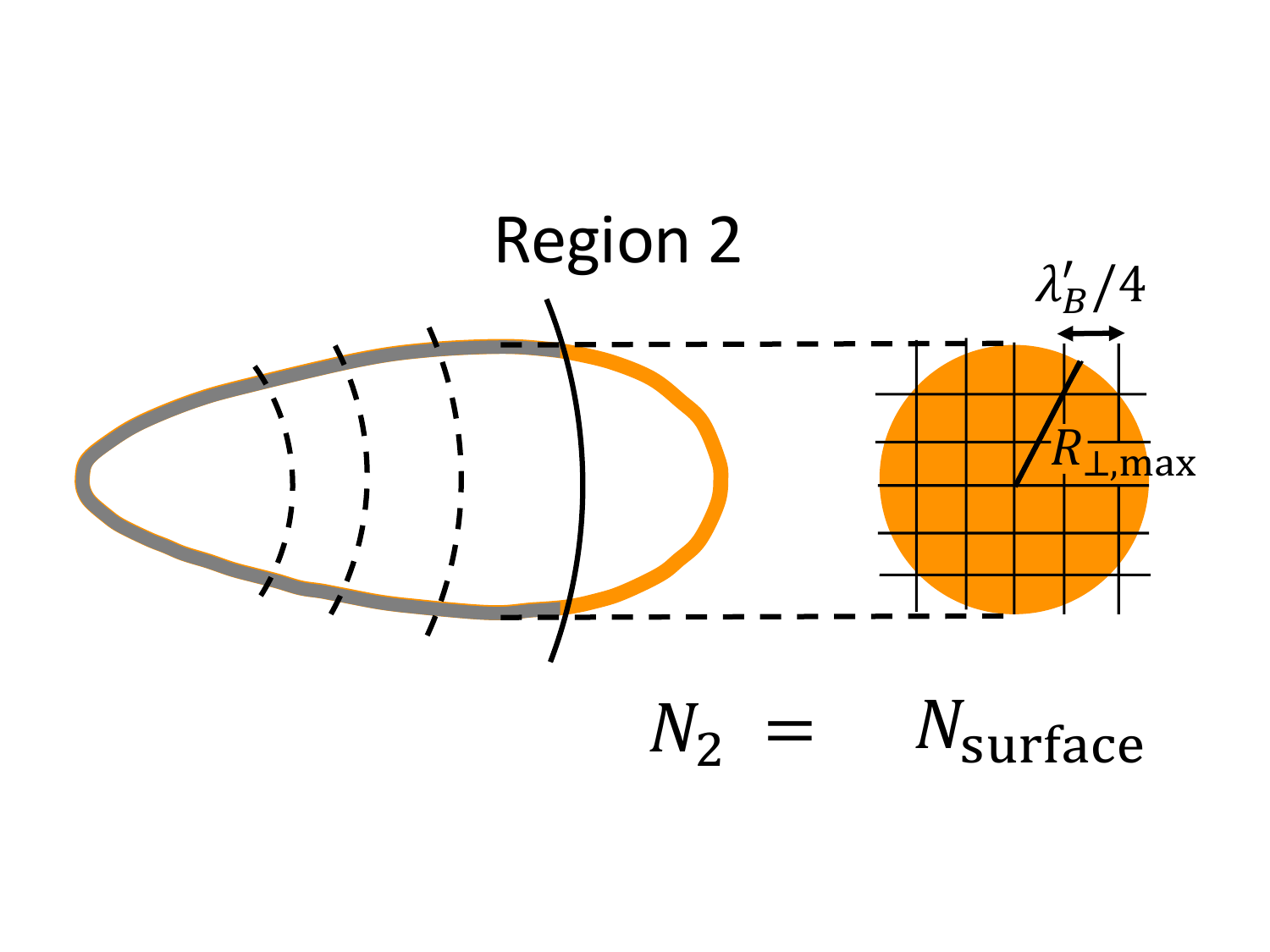}
    \caption{Schematic pictures of the EATS and their images viewed from the observer with $\theta_v = 0$ at the optical band (left) and at the radio band (right) before the jet break time. The orange region is the brightest part of the EATS, which is called Region 1 at the optical band and Region 2 at the radio band.} 
    \label{fig:patch}
\end{figure*}

We analytically estimate the level of PDs before showing our numerical results in \Secref{sec:results}. 
By this analysis we can understand the order of magnitude of PD at each band and its dependence on $f_B$.

The average level of PD with the hydrodynamic-scale turbulent magnetic field can be estimated by 
\begin{equation}
    \Pi_\nu \sim \frac{\Pi_0}{\sqrt{N}}, \label{eq:PD-largescale}
\end{equation}
where $N (\gg 1)$ is the number of patches with coherent magnetic field in the visible area of the shocked region \citep{Gruzinov1999}, as introduced in \Secref{sec:intro}. 
To estimate $N$ at the optical and radio bands,
we divide the EATS into three parts; the dark region at both of the wavelengths ($y \lesssim 0.4$, Region 0), the bright region at optical ($y \sim 0.5-0.7$, Region 1) and the bright region at radio ($y \sim 0.8-1.0$, Region 2), as shown by the schematic pictures in \Figref{fig:patch}. 
The net PD is the brightness weighted average of local PDs. Thus we may estimate the optical and radio PDs by counting $N$ of Region 1 ($N_1$) and Region 2 ($N_2$), respectively.

The distance of a position on the EATS from the line of sight is $R_\perp \simeq (\sqrt{2}R_l/4\gamma_l)\sqrt{y-y^5}$, and its maximum value is given at $y \simeq 0.67$ as $R_{\perp,\rm max} \simeq 0.26 R_l/\gamma_{l}$. 
Region 1 can be regarded as a cylindrical surface of the radius $R_{\perp,{\rm max}}$. In this cylindrical surface the number of parts between shocks with different configurations of turbulent magnetic field is
\begin{equation}
    N_f \sim \frac{R}{\delta R_B} = 4 f_B^{-1}.
\end{equation}
The number of patches in each of the parts is
\begin{equation}
    N_\phi = \frac{2\pi R_{\perp,\rm max}}{\lambda'_{B}/4} \simeq 380 f_B^{-1},
\end{equation}
where we have assumed that the size of a patch is a quarter of wavelength of turbulent magnetic field, i.e., $\sim \lambda'_{B}/4$, and estimated $\lambda'_B$ at $y=0.6$. 
We obtain $N_1 = N_\phi N_f \sim 1500 f_B^{-2}$, and then we have analytical estimate of optical PD as
\begin{equation}
    \Pi_{\rm a,opt} \sim \frac{\Pi_0(\nu \geq \nu_m)}{\sqrt{N_1}} \sim 2 f_B \;\%.
    \label{eq:PD-Region 1}
\end{equation}

Region 2 can be roughly regarded as a spherical surface of the radius $R_{\perp,{\rm max}}$.
The number of patches in this surface $N_{\rm surface}$ is
 \begin{equation}
     N_{\rm surface} \simeq \frac{\pi R_{\perp,\rm max}^2}{(\lambda'_{B}/4)^2} \simeq 880 f_B^{-2}, \label{eq:N-surface-thinshell}
 \end{equation}
where we have estimated $\lambda'_B$ at $y=1.0$.
We obtain $N_2 \sim N_{\rm surface}$, and then we have analytical estimate of radio PD as
\begin{equation}
    \Pi_{\rm a,radio} \sim \frac{\Pi_0(\nu < \nu_m)}{\sqrt{N_2}} \sim 2 f_B\; \%.
    \label{eq:PD-Region 2}
\end{equation}

From \Eqref{eq:PD-Region 1} and (\ref{eq:PD-Region 2}), our analytical estimate shows that $\Pi_{\rm a,opt}$ and $\Pi_{\rm a,radio}$ are comparable and both proportional to $f_B$.

\begin{figure*}[htbp]
    \begin{minipage}[c]{0.5\linewidth}
        \centering
        \includegraphics[width=8.2cm]{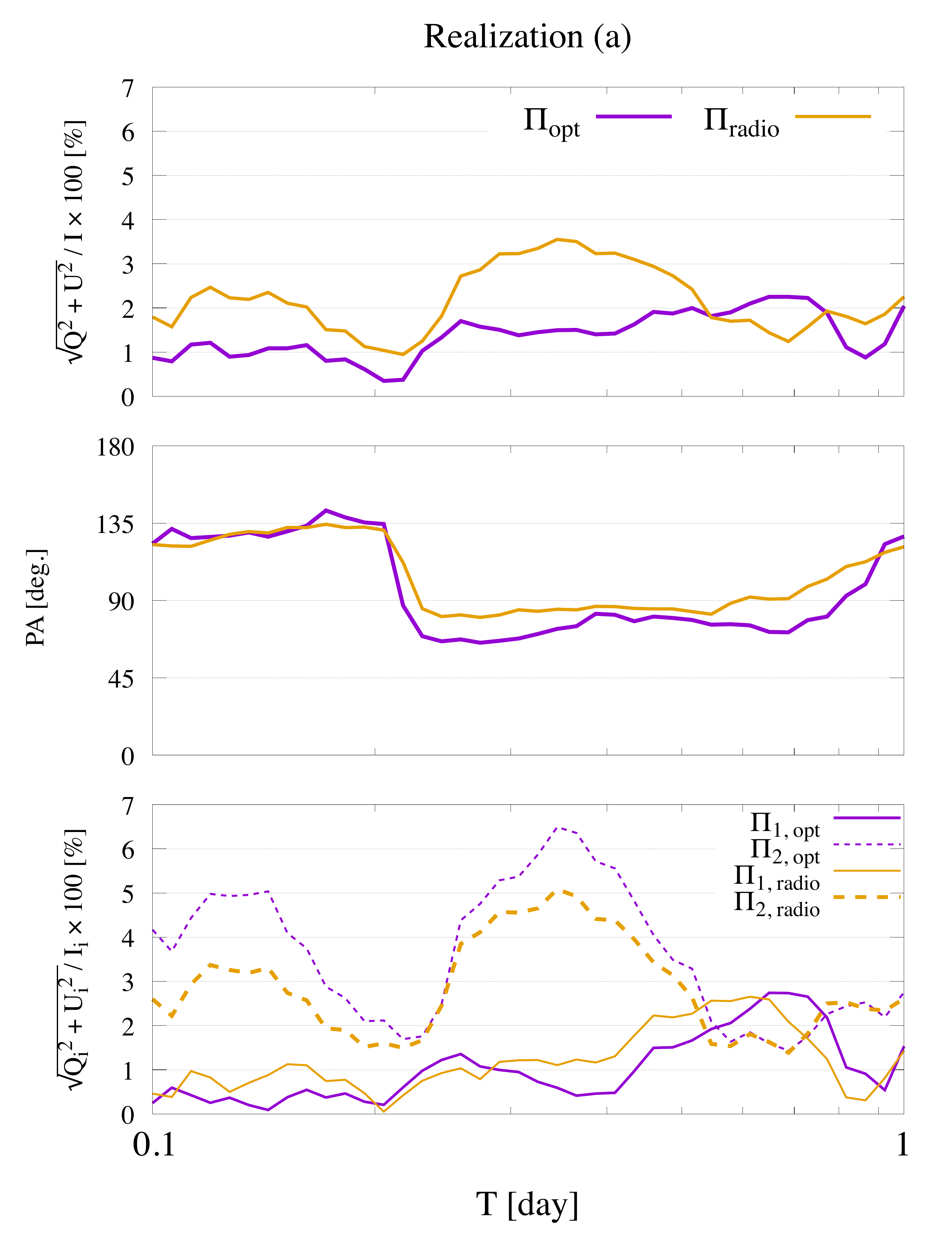}
    \end{minipage}
    \begin{minipage}[c]{0.5\linewidth}
        \centering
        \includegraphics[width=8.2cm]{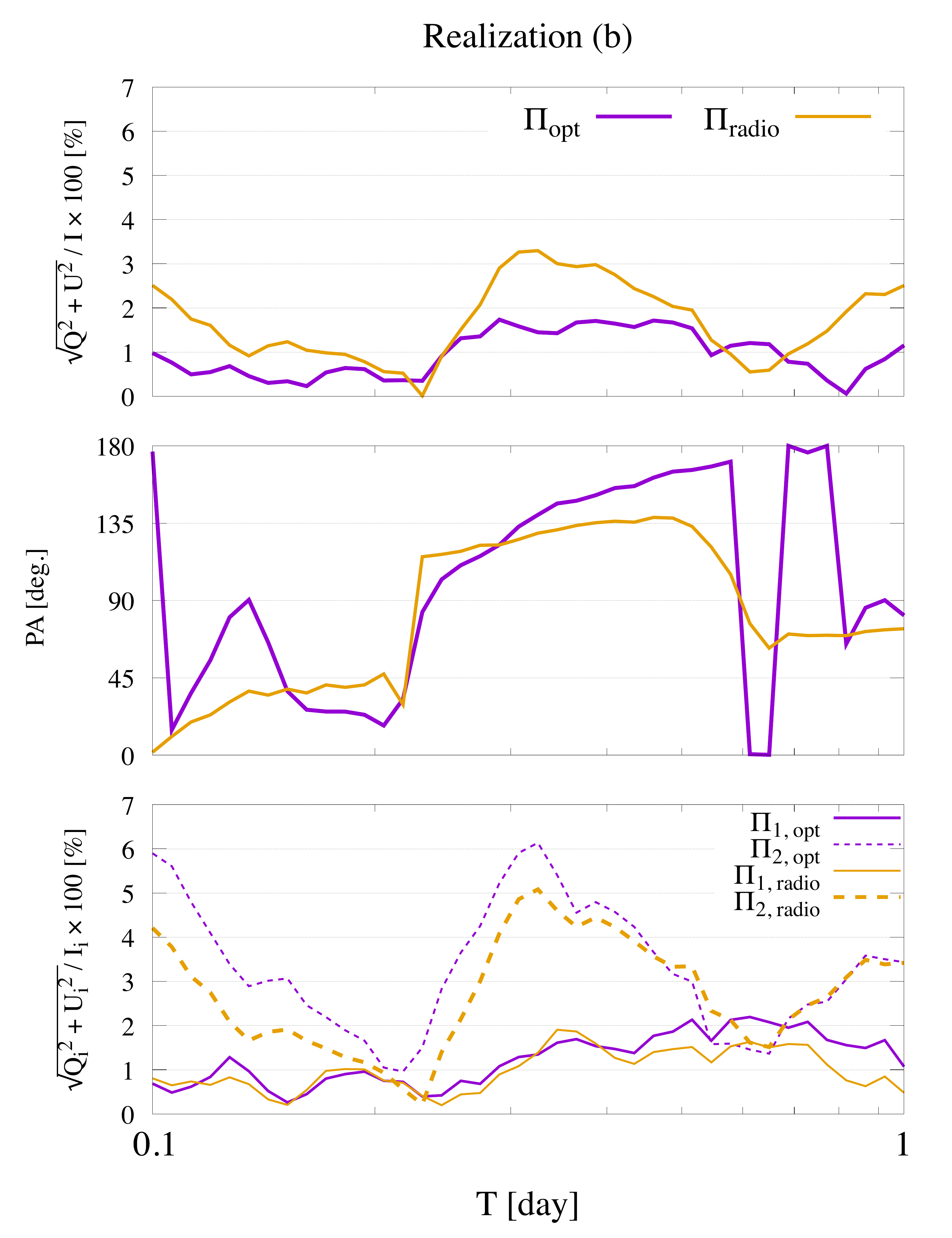}
    \end{minipage}\\
    \begin{minipage}[c]{0.5\linewidth}
        \centering
        \includegraphics[width=8.2cm]{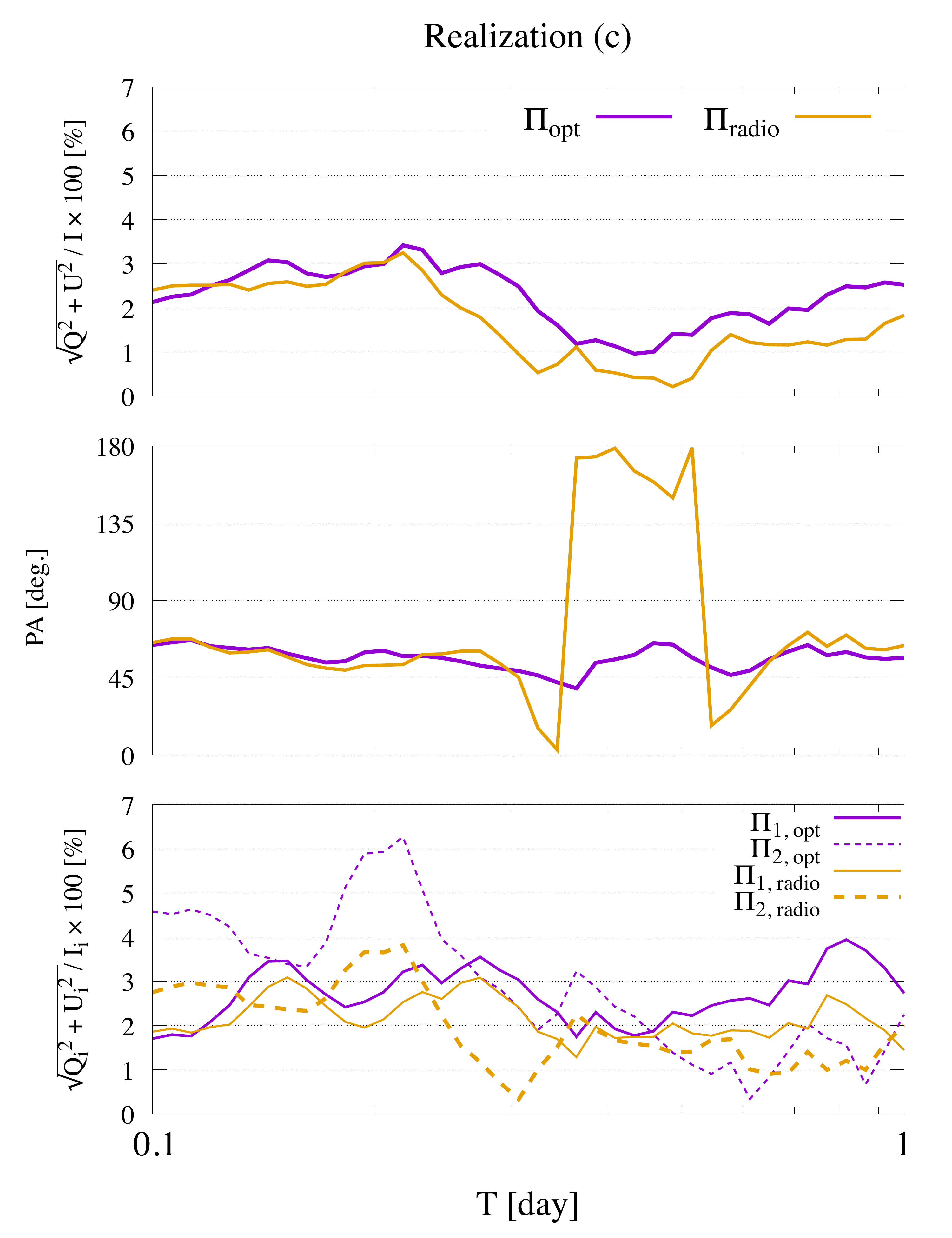}
    \end{minipage}
    \begin{minipage}[c]{0.5\linewidth}
        \centering
        \includegraphics[width=8.2cm]{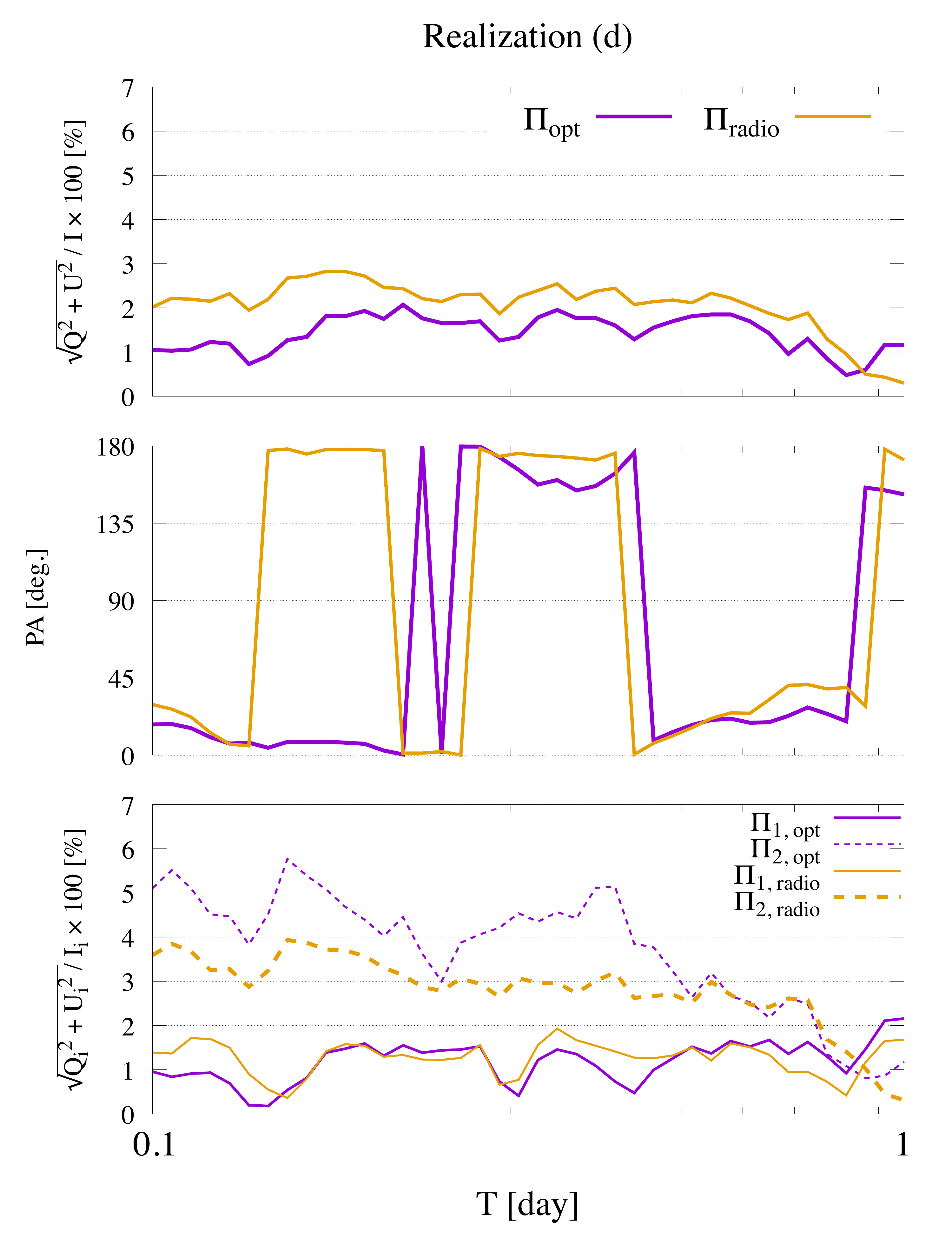}
    \end{minipage}
    \caption{
    PDs and PAs as functions of time $T$ for $f_B = 1.0$ at frequencies $10^{15}$ Hz  (optical) and $100$ GHz (radio) during $T = 0.1-1.0~{\rm days}$.
    Realizations (a), (b), (c), and (d) are the calculation results for different realizations of turbulent magnetic field. The top, middle, and bottom panels represent the net PD curves, the net PA curves, and the PD curves at Region 1 ($0.42 \le y < 0.76$; solid lines) and Region 2 ($y \ge 0.76$; dashed lines), respectively (purple lines for optical and orange lines for radio).
    \label{fig:PDPA-curve}}
\end{figure*}

\section{Numerical Calculation Results} \label{sec:results}
We show the numerical calculation results of synchrotron polarization from collimated blast waves as functions of $T$ and its dependence on $f_B$ in this section.

\subsection{Numerical setup}
For our numerical calculations, we set the number of waves to be $N_m = 3000$ in \Eqref{eq:B-field}. From analysis of emission from a plane of uniform brightness with various values of $N_m$, we found that the net PD is converged to be $\sim \Pi_0/\sqrt{N}$ for $N_m \gg N$. If $N_m$ is too small, PD induced by an artificial anisotropy of turbulent field $(> \Pi_0/\sqrt{N})$ arises. 
We confirmed $N_m \gg N$ for our calculations, so that our resulting PDs are not affected by the numerical artifact.

\begin{figure}[t]
    \raggedleft
    \includegraphics[width=8.5cm]{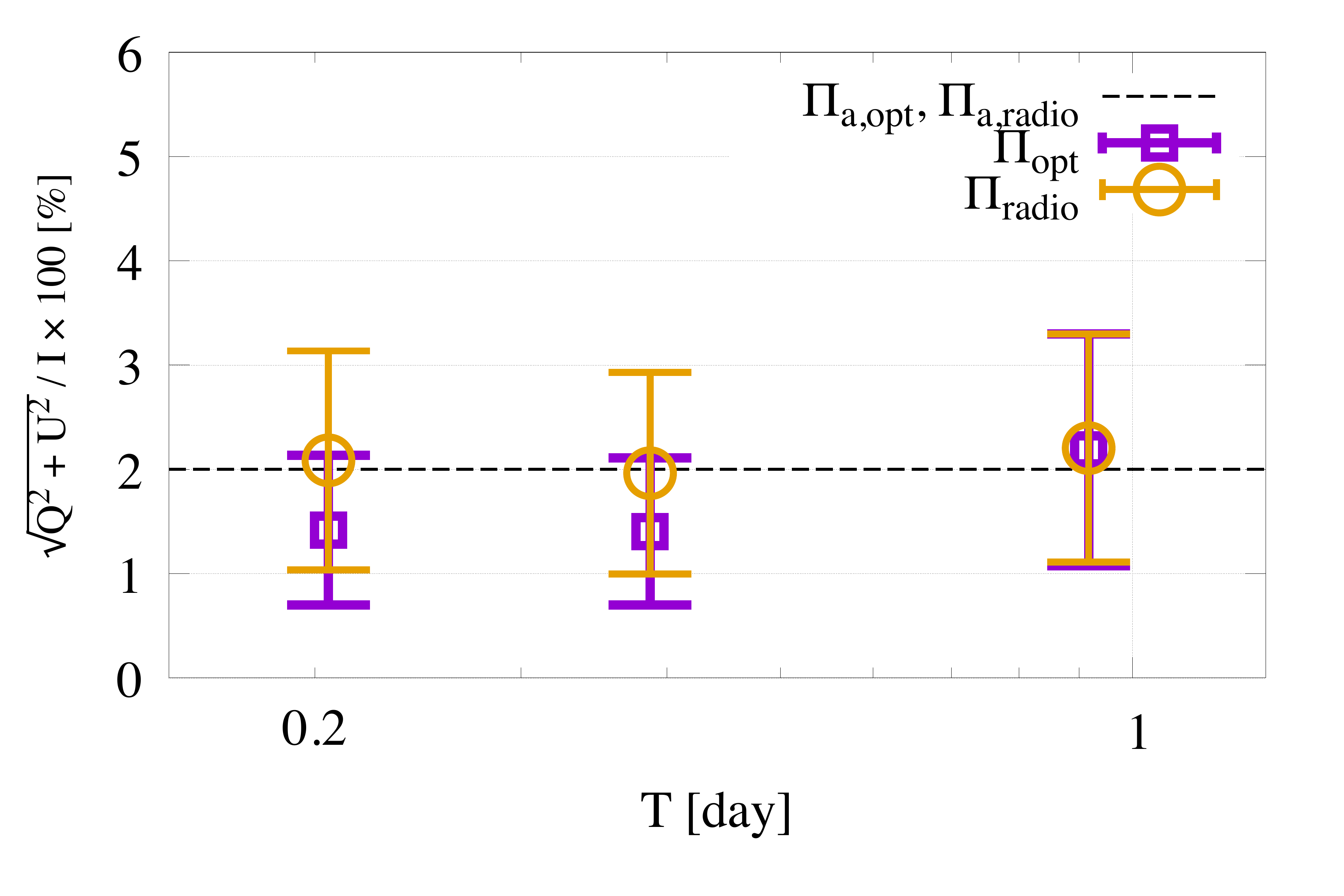}
    \caption{Averaged PDs (the purple square for optical and the orange circle for radio) and the standard deviations (the error bars) for 300 turbulent field realizations at $T=0.2,\,0.4$ and $0.9$ days for $f_B = 1.0$. 
    The horizontal dashed lines represent the analytically estimated optical and radio PD (\Eqref{eq:PD-Region 1}, (\ref{eq:PD-Region 2})).}
    \label{fig:PD-time}
\end{figure}

We set the number of grids of the spherical thin shell for giving the magnetic field as $(60/f_B,~240/f_B)$ in $(R,\phi)$ directions. We set the number of the spatial grids for the flux and polarization calculations as $(256, 1024)$ in $(y,\phi)$ directions. We confirmed the numerical convergences of PD and PA at every observed time with the above setups.

\subsection{Temporal behaviors of PDs} \label{subsec:PD-curve}
\Figref{fig:PDPA-curve} shows the calculated PD curves (the first panels) and PA curves (the second panels) for $f_B = 1.0$ at frequencies $10^{15}$ Hz (optical), $100$ GHz (radio) during $T = 0.1-1.0$ days. We show the calculation results for four cases, Realizations (a), (b), (c), and (d), for which the parameter sets are the same as listed in Table 1, but the realizations of random numbers for the turbulent field creation are different. Interestingly, we have the optical PD of the observed level, $\Pi_{\rm opt} \simeq 1-3\%$, in the case of $f_B = 1.0$. 

The temporal variations of PDs and PAs look random and continuous at both of the optical and radio bands. This is a behavior different from that in the plasma-scale turbulent field model, in which PDs have one or two peaks, and PAs keep constant or have a sudden flip by $90\;$deg \citep{Sari1999,Ghisellini1999,Rossi2004}. We also find that for some realizations, the time interval in which radio PDs are higher than the optical PDs is frequently seen (see Realizations (a) and (b)) while at other realizations, the opposite trend is seen (see Realization (c)).
The radio PD higher than the optical one is a distinct feature from the plasma-scale turbulent field model, in which the radio PD is always lower than the optical one \citep{ST2021,Rossi2004}. 

As we mentioned in the above sections, the difference between the radio and the optical polarizations comes from the frequency dependence of the brightness profile of EATS
shown in \Figref{fig:flux-density-element}. 
To demonstrate this, we numerically calculate PDs both at Region 1 and 2 by
\begin{equation}
    \Pi_{i, \nu} = \frac{\sqrt{Q^2_{i,\nu} + U^2_{i,\nu}}}{I_{i,\nu}} \quad (i=1,2), \label{eq:PD1_PD2}
\end{equation}
where the subscript $i$ denotes a region number and $I_{i,\nu}, Q_{i,\nu},$ and $U_{i,\nu}$ are the Stokes parameters calculated at each region (cf. \Eqref{eq:I-nu}-(\ref{eq:U-nu})). Here, we set the ranges of Regions 0, 1, and 2 as $y < 0.42$, $0.42 \le y < 0.76$, $y \ge 0.76$, respectively. We have chosen $y = 0.76$ (and at $y = 0.42$ at the optical band) as the point where the value of $dF/dy$ is 60\% of its maximum at each band (see also \Secref{subsec:validity-thin-shell}).
We show these calculated PDs in the bottom panels of \Figref{fig:PDPA-curve}. 
We can see that $\Pi_{\rm radio} > \Pi_{\rm opt}$ while $\Pi_{2,{\rm radio}} > \Pi_{1,{\rm opt}}$, and vice versa. This means that the net PD reflects the PD at the brightest region at each band. It is remarkable in Realizations (a) and (b) that the temporal behaviors of $\Pi_{\rm radio}$ are in good agreement with $\Pi_{2,{\rm radio}}$, not with $\Pi_{1,{\rm radio}}$.

\begin{figure}[t]
    \raggedleft
    \includegraphics[width=8.5cm]{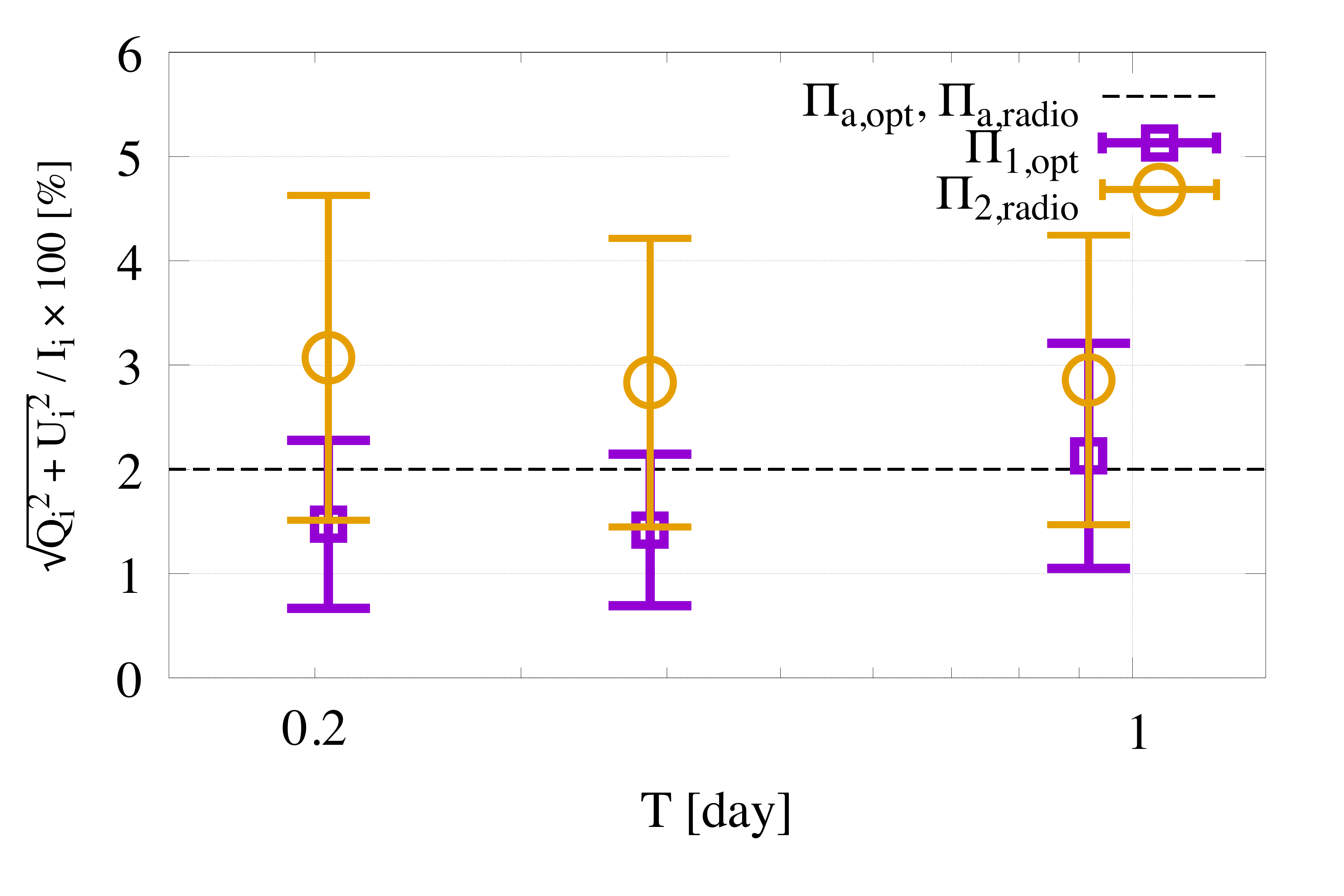}
    \caption{Same as \Figref{fig:PD-time}, but PDs are calculated at Region 1 at the optical band (the purple square marks with error bars) and at Region 2 at the radio band (the orange circle marks with error bars), respectively.}
    \label{fig:PD-time-region}
\end{figure}

\begin{figure}[t]
    \centering
    \includegraphics[width=8.5cm]{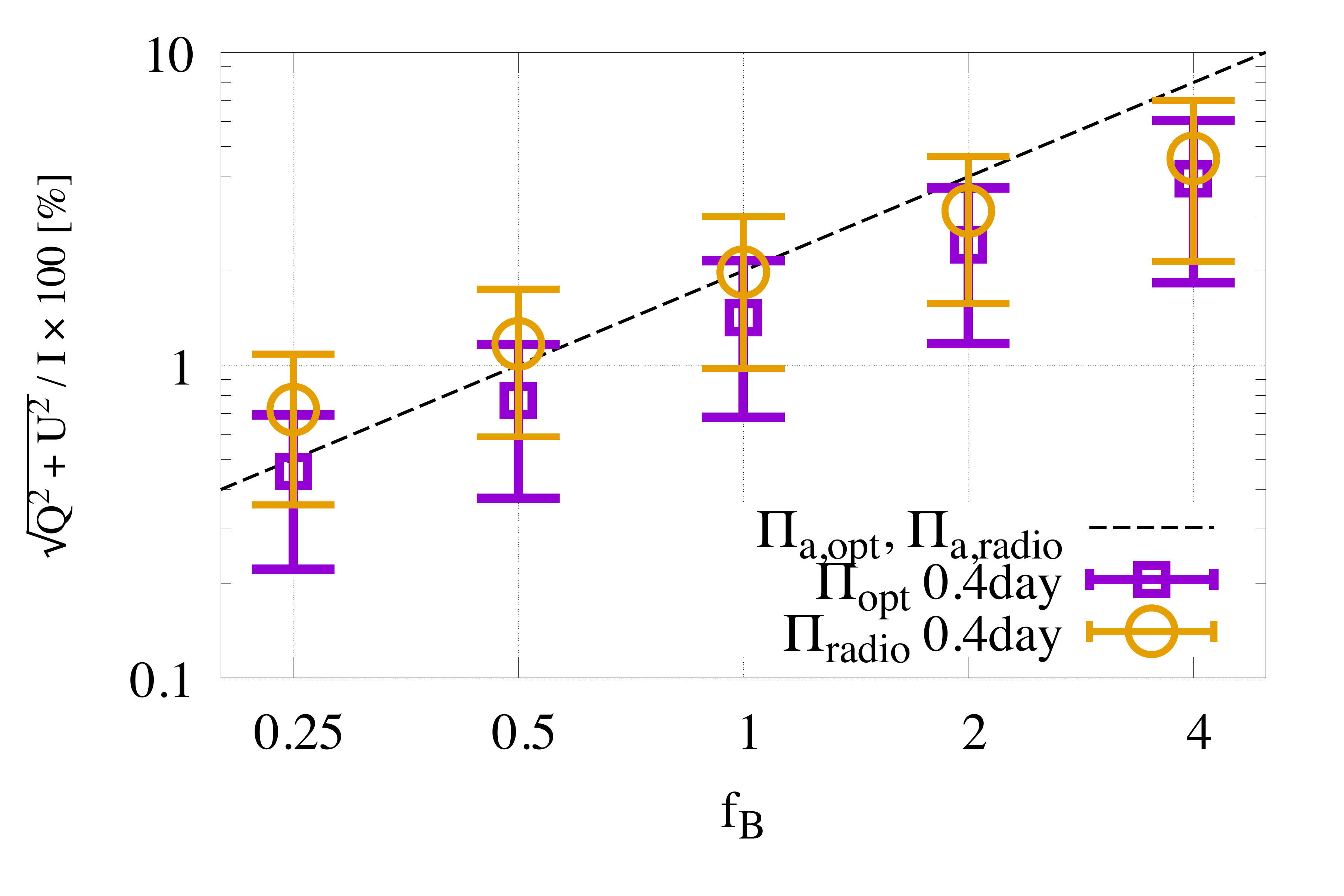}
    \caption{Realization-averaged PDs at $T=0.4~{\rm days}$ for $f_B = 0.25,\,0.5,\,1.0,\,2.0,$ and $4.0$. The marks are the same as \Figref{fig:PD-time}. The dashed line is analytically estimated optical PD and radio PD (\Eqref{eq:PD-Region 1}, (\ref{eq:PD-Region 2})). \label{fig:fB-PD}}
\end{figure}

\subsection{Statistical behavior of PDs} \label{subsec:stat-PD}
We calculate the PDs using the magnetic field turbulence for 300 turbulence realizations, and take average and variance. We perform this calculation for $f_B=1.0$ and show the results at $T=0.2,\,0.4$ and $0.9\,$days in \Figref{fig:PD-time}.
At $T=0.2,\,0.4\,{\rm days}~(<T_j)$, the realization-averaged PDs at the radio band $\langle \Pi_{\rm radio} \rangle$  (the orange circle marks) tend to be higher than that at the optical band $\langle \Pi_{\rm opt} \rangle$ (the purple square marks). On the other hand, at $T=0.9\; {\rm day}~(>T_{j})$, $\langle \Pi_{\rm radio} \rangle$ is comparable to $\langle \Pi_{\rm opt} \rangle$.
This is because the part of the EATS that would be the brightest at the optical band is outside the jet and does not radiate (see \Figref{fig:flux-density-element}), and then the brightness profile becomes more alike that at the radio band. 

In \Figref{fig:PD-time-region}, we show the realization-averaged PDs at Region 1 at the optical band $\langle \Pi_{\rm 1,opt} \rangle$ and that at Region 2 at the radio band $\langle \Pi_{\rm 2,radio} \rangle$. While our rough estimate leads to $\Pi_{\rm a,radio} \sim \Pi_{\rm a,opt}$ (see \Secref{sec:analytical-estimate}), the numerical results show that $\langle \Pi_{\rm 2,radio} \rangle$ is a bit higher than $\langle \Pi_{\rm 1,opt} \rangle$. One can see that this trend results in $\langle \Pi_{\rm radio} \rangle \gtrsim \langle \Pi_{\rm opt} \rangle$ shown in \Figref{fig:PD-time}. \Figref{fig:PD-time-region} also shows a similar time evolution as \Figref{fig:PD-time}. $\langle \Pi_{\rm 2,radio} \rangle$ is higher than $\langle \Pi_{\rm 1,opt} \rangle$ at $T < T_j$, whereas at $T = 0.9 \;{\rm day}~(> T_j)$, $\langle \Pi_{\rm 1,opt} \rangle$ slightly increases and closer to $\langle \Pi_{\rm 2,radio} \rangle$.

In \Figref{fig:fB-PD}, we show the PDs at $T = 0.4$ days for various values of $f_B = 0.25 - 4.0$. The PDs are roughly proportional to $f_B$. In \Figref{fig:PD-time}-\ref{fig:fB-PD}, we also show the lines of $\Pi_{\rm a,opt}$ and $\Pi_{\rm a,radio}$ (see \Secref{sec:analytical-estimate}), which well explain the order of magnitudes of PDs and the dependence of PDs on $f_B$ except for $f_B=4.0$. Note that at $f_B=4.0$, the analytical estimate may be invalid, because \Eqref{eq:PD-largescale} is an expression in the large $N$ limit.

\section{Summary and Discussion} \label{sec:summary}

We have constructed a semi-analytic model of GRB forward shock afterglows with hydrodynamic-scale turbulent magnetic field and performed analytical and numerical estimates of optical and radio polarizations. 
The numerical calculations are under the assumption of the zero viewing angle and the isotropic turbulent magnetic field.
Our analytical estimate calculating the number of coherent field patches in the brightest part of the EATS at each frequency band indicates the comparable level of optical and radio PDs $\sim 2 f_B \%$. Interestingly, the observed level $\sim 1-3 \%$ of late-time optical GRB afterglows is obtained for $f_B \sim 1$.
The numerical calculations are consistent with the analytical estimate, but show that the realization-averaged PDs at the radio band $\langle \Pi_{\rm radio} \rangle$ are slightly higher than that at the optical band $\langle \Pi_{\rm opt} \rangle$ at $T < T_j$. 
We also numerically show that the averaged radio PD at Region 2 $\langle \Pi_{2,\rm radio}\rangle$ is a bit higher than the optical PD at Region 1 $\langle \Pi_{1,\rm opt} \rangle$ at $T < T_j$, which causes $\langle \Pi_{\rm radio} \rangle \gtrsim \langle \Pi_{\rm opt} \rangle$, and that $\langle \Pi_{2,\rm radio}\rangle$ and $\langle \Pi_{1,\rm opt} \rangle$
exhibit similar time evolution as $\langle \Pi_{\rm radio} \rangle$ and $\langle \Pi_{\rm opt} \rangle$, respectively.

Our numerical calculations also show that PDs and PAs vary randomly and continuously at both of two bands. 
In some time intervals, the radio PD can be significantly higher than the optical PD.
In contrast, in the plasma-scale turbulent magnetic field model, (i) the radio PD is always lower than the optical PD, (ii) PD have one or two peaks while PA is constant or has sudden flip by $90^\circ$,
and (iii) the differences in PAs between the radio and the optical bands is zero or $90^\circ$ \citep{Sari1999,Ghisellini1999,Rossi2004,ST2021}. 
Thus, the more simultaneous polarimetric observations of GRB afterglows at the radio and the optical bands would be decisive tests of these two turbulent magnetic field models.

\subsection{the validity of the thin-shell approximation} \label{subsec:validity-thin-shell}
\begin{figure}[t]
    \centering
    \includegraphics[width=8.5cm]{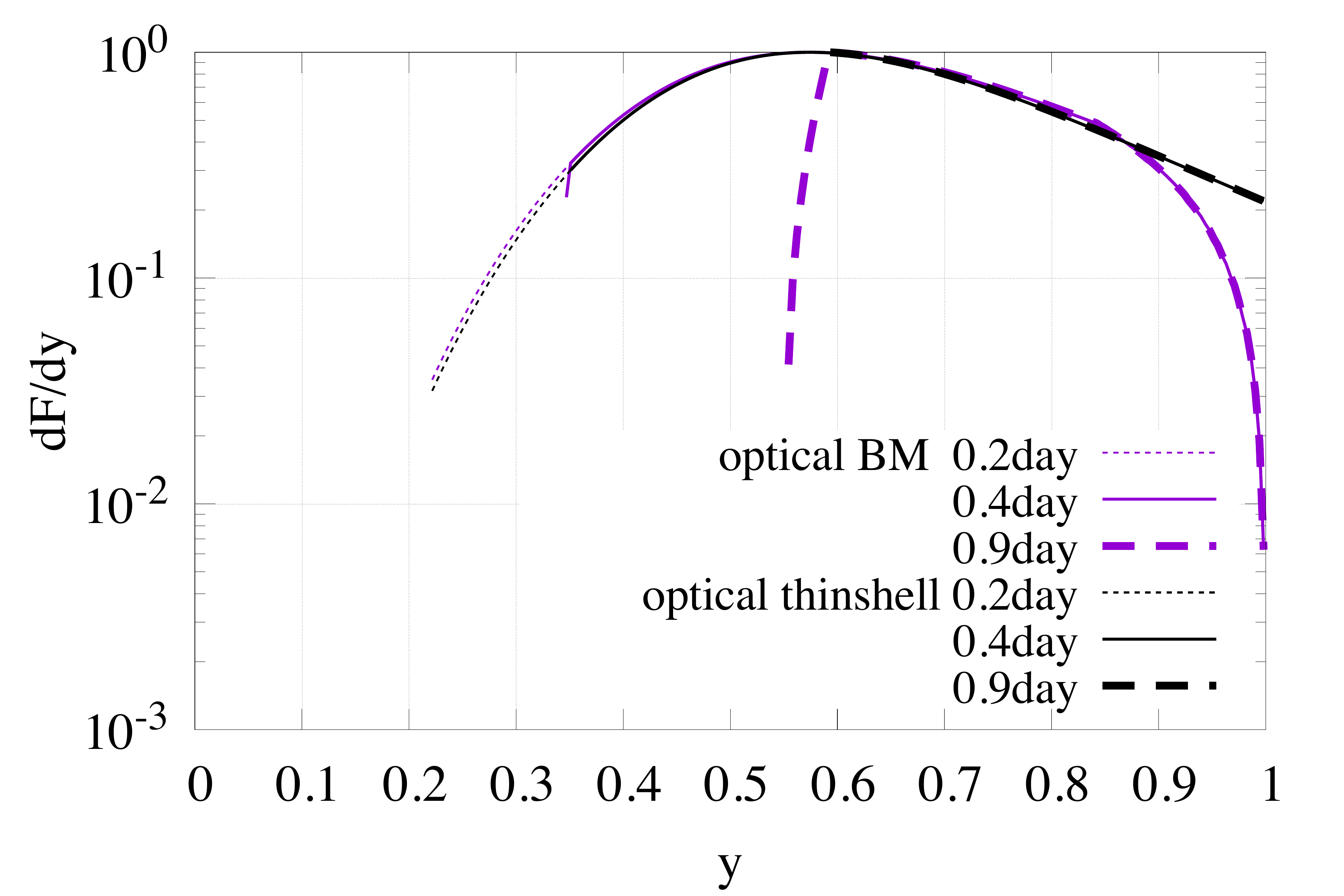}
    \includegraphics[width=8.5cm]{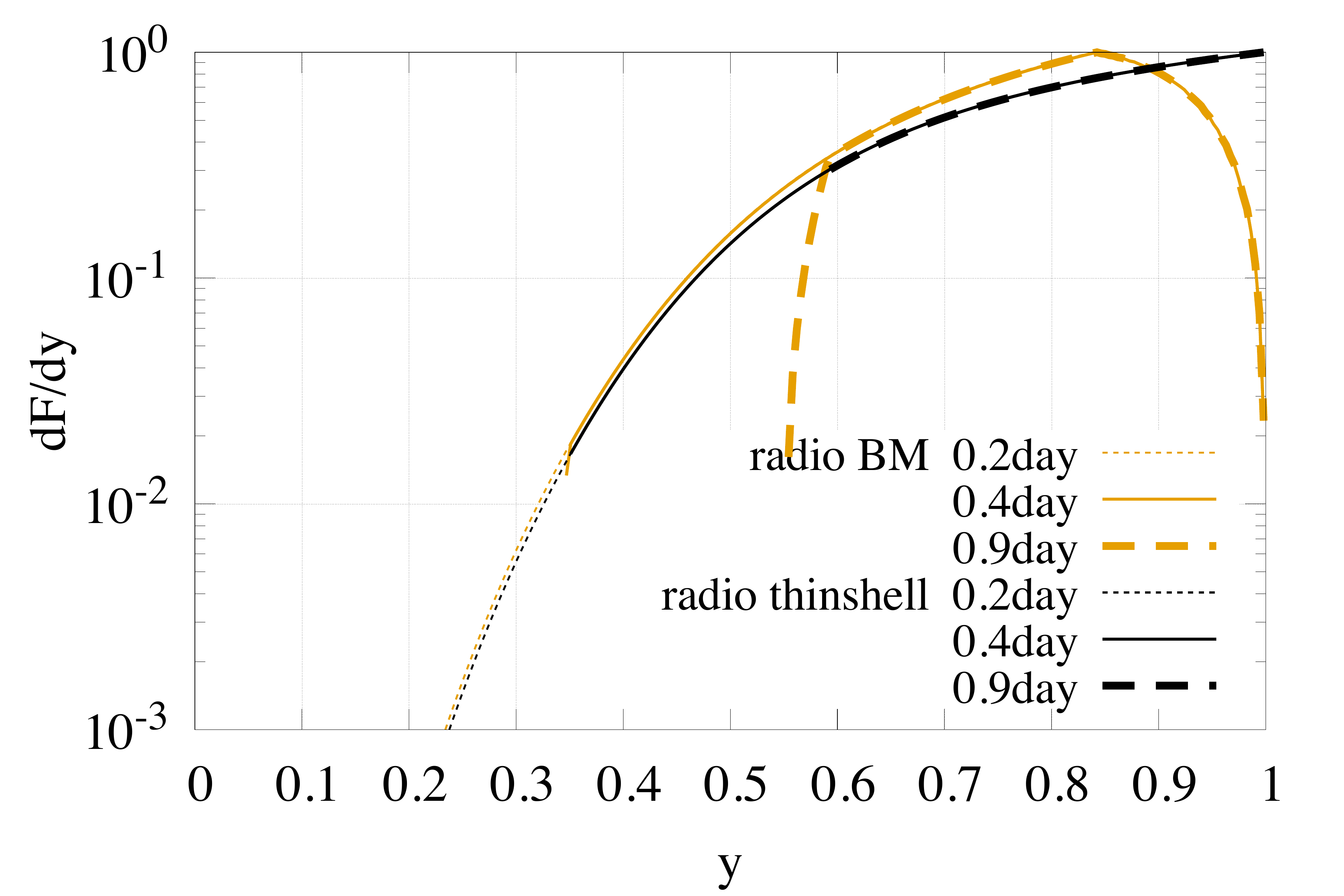}
    \caption{Top panel: The $\phi, \chi$-integrated flux density element $dF/dy$ (in arbitrary units) using the BM structure for the fluid behind the shock front (purple) and thin-shell approximation (black) at the optical band at $T = 0.2$ (dotted lines), 0.4 (solid lines), 0.9 (dashed lines) days. Note that for the BM structure, the $\chi$-integral interval is $\chi \leq 2$. Bottom panel: Same as the top panel, but at the radio band. The BM structure is represented by the orange lines. \label{fig:flux-density-element-BM}}
\end{figure}

We have used the thin-shell approximation for the shocked region. Here we discuss its validity. 
If we use the BM structure for the fluid behind the shock front, the brightness profile is changed from that of the thin-shell approximation, which could affect the frequency dependence of PD.
The flux density element of the BM structure is $dF/d\chi dy$ (after the integration over $\phi$), where $\chi$ is a self-similar variable corresponding to the distance behind the shock front \citep[cf. Eq. (9) of][]{Granot1999a}. 
For the thin-shell approximation, we have roughly counted the number of coherent patches $N$ on the region of EATS with high value of $dF/dy$, by which we have obtained the net PD consistent with the numerical results (see \Secref{sec:analytical-estimate}). For the BM structure, $N$ can be counted roughly by deriving the profile of $dF/dy$ after the integration over $\chi$, recalling that the magnetic field structure behind the shock front is assumed to be unchanged while the shock expands over the distance of $\delta R_B$.
We integrate $dF/d\chi dy$ in the range of $\chi = 1-2$, because only this region is bright and mainly contributes to the net PD, and plot it in \Figref{fig:flux-density-element-BM}, where $dF/dy$ for the thin-shell approximation is also shown for comparison.

At the optical band, as shown in the top panel of \Figref{fig:flux-density-element-BM}, the values of $dF/dy$ for the BM structure and for the thin-shell approximation are the same except at $y>0.9$ and take the same maximum value at $y \sim 0.6$. Thus, we suppose that our calculated $\Pi_{\rm opt}$ will not be changed if we calculated that with the BM structure. 

At the radio band, as shown in the bottom panel of \Figref{fig:flux-density-element-BM}), $dF/dy$ for the BM structure peaks at $y\sim 0.85$ while that with the thin-shell approximation peaks at $y=1.0$. Then we should estimate the number of coherent patches $N$ for the BM structure. The brightest region for the BM structure at the radio band has a ring-like shape with the inner radius $R_\perp~(y=0.85)$ and the outer radius $R_{\perp, \rm max})$. The number of patches in this ring $N_{2,\rm BM}$ is 
\begin{equation}
    N_{2,\rm BM} = \frac{\pi(R_{\perp, \rm max}^2 - R_\perp^2(y=0.85))}{(\lambda'_B(y=0.85)/4)^2} \times 2 \simeq 920 f_B^{-2},
\end{equation}
where the last factor $2$ corresponds to $N_f$ introduced in \Secref{sec:analytical-estimate} and means that the region between $y=0.67-0.85$ includes at most two different configurations of turbulent magnetic field.
$N_{2,\rm BM}$ is comparable to that with the thin-shell approximation (see \Eqref{eq:N-surface-thinshell}). Therefore, if we calculate $\Pi_{\rm radio}$ with the BM structure, it would not significantly change from our results with the thin-shell approximation.

In summary, our conclusion on $\Pi_{\rm radio}$ and $\Pi_{\rm opt}$ in this paper would not be changed even if we calculate the polarization with the BM structure.

\subsection{future work}
Our analysis treats only the case of zero viewing angle $(\theta_v = 0)$ and isotropic turbulent magnetic field $({\sigma'}^2_\parallel = {\sigma'}^2_\perp/2)$ in this paper. In reality, many GRBs should be observed for finite values of $\theta_v$. The shock jump condition would make the magnetic field anisotropic in the downstream region even if it is isotropic in the upstream region \citep{Ma&Zhang2022}. The nonlinear dynamics of MHD turbulence also show anisotropic magnetic energy spectrum \citep{Goldreich1995, Lazarian1999}. The cases of finite $\theta_v$ (particularly $0 < \theta_v < \theta_j$, for which the prompt emission can be bright) and/or anisotropic turbulent field will exhibit different features of PDs and PAs from those we have found in this paper, which are interesting theoretically and observationally. The effects of ordered field component and Faraday effects should also be examined \citep[cf.][]{Granot2003, Toma2008}. We will perform calculations in those cases for $T \gtrsim 1\;$day and compare the results with the observed data in separate papers. 
This will enable us to distinguish the hydrodynamic-scale turbulent field and the plasma-scale turbulent field in the regions behind GRB afterglow forward shocks.

\vskip\baselineskip
We thank S. Chon for helpful discussions. We also thank the anonymous referee for useful comments.
Numerical calculations were performed on Draco, a computer cluster of FRIS. We utilized Science Lounge of FRIS CoRE for discussions many times. This work is partly supported by JSPS Grants-in-Aid for Scientific Research No. 18H01245 (K.T.), No. 22K14028 (S.S.K.), by Graduate Program on Physics for the Universe (GP-PU), Tohoku University (A.K.), and by the Tohoku Initiative for Fostering Global Researchers for Interdisciplinary Sciences (TI-FRIS) of MEXT's Strategic Professional Development Program for Young Researchers (S.S.K.).

\appendix
\renewcommand{\thetable}{\Alph{section}.\arabic{figure}}
\renewcommand{\thefigure}{\Alph{section}.\arabic{table}}

\section{polarization from Kolmogorov-type turbulent magnetic field} \label{app:Kolmogorov-spectrum}
We have constructed the turbulent magnetic field model with wavelength of $\lambda'_B = f_B \Delta R'$ for the shocked region of the forward shocks in \Secref{sec:large-B} and showed the calculation results of synchrotron polarization in \Secref{sec:results}. Here we consider the distribution of magnetic field wavelengths. The MHD simulations of Richtmyer-Meshkov instability at relativistic shocks show a magnetic field turbulence with power spectrum somewhat flatter than Kolmogorov one \citep{InoueT2011, Mizuno2014}. 
Observations of supernova remnants indicate the Kolmogorov power spectra in the shocked regions \citep{Shimoda2018, Shimoda2022, Vishwakarma2020}. 
We assume isotropic turbulent field (i.e., ${\sigma'}^2_\parallel = {\sigma'}^2_\perp/2$) with a Kolmogorov power spectrum as an example just to demonstrate the effect of field wavelength distribution on the synchrotron polarization.

The Kolmogorov power spectrum is $P(k)dk \propto k^{-11/3} 4\pi k^2 dk$, i.e. $P(k)k d\ln k \propto k^{-2/3} d\ln k$. We set $\sigma_{\perp}^2, \sigma_{\parallel}^2 \propto P(k)k$ in \Eqref{eq:B-field-direction} for $k_{\rm min} \leq k \leq k_{\rm max}$, where $k_{\rm min} = 2\pi/ \lambda'_B$. 
We calculate the polarization for $x \equiv k_{\rm max}/k_{\min} = 2, 4,$ and $8$. For the calculation of $x=8$, we have divided the $\ln k$-space by 8 grids and set $3000$ waves of random wave directions and phases for each grid. Then we find that the obtained PDs are roughly proportional to a characteristic length,
\begin{equation}
    L_c = \frac{\int_{k_{\rm min}}^{k_{\rm max}} \frac{2\pi}{k} P(k) dk}{\int_{k_{\rm min}}^{k_{\rm max}} P(k) dk} = \frac{4 \pi}{5}k_{\rm min}^{-1} \frac{x^{5/3}-1}{x(x^{2/3}-1)}. 
    \label{eq:coherent-length}
\end{equation}
$L_c(x)$ monotonically decreases and converges to $(4\pi/5) k_{\rm min}^{-1}$.
Although the power spectrum will extend to the tiny dissipation scale (i.e., $x \gg 1$) which depends on parameters such as particle mean free path, viscosity, resistivity, etc. \citep{Schekochihin2002}, we do not have to set such a large value of $x$. 
Since $L_c(x=8) \simeq L_c(x\gg 1)$, the calculation of polarization for $x = 8$ is a good approximation of that for $x \gg 1$.

\begin{figure}[ht]
    \centering
    \includegraphics[width=8.5cm]{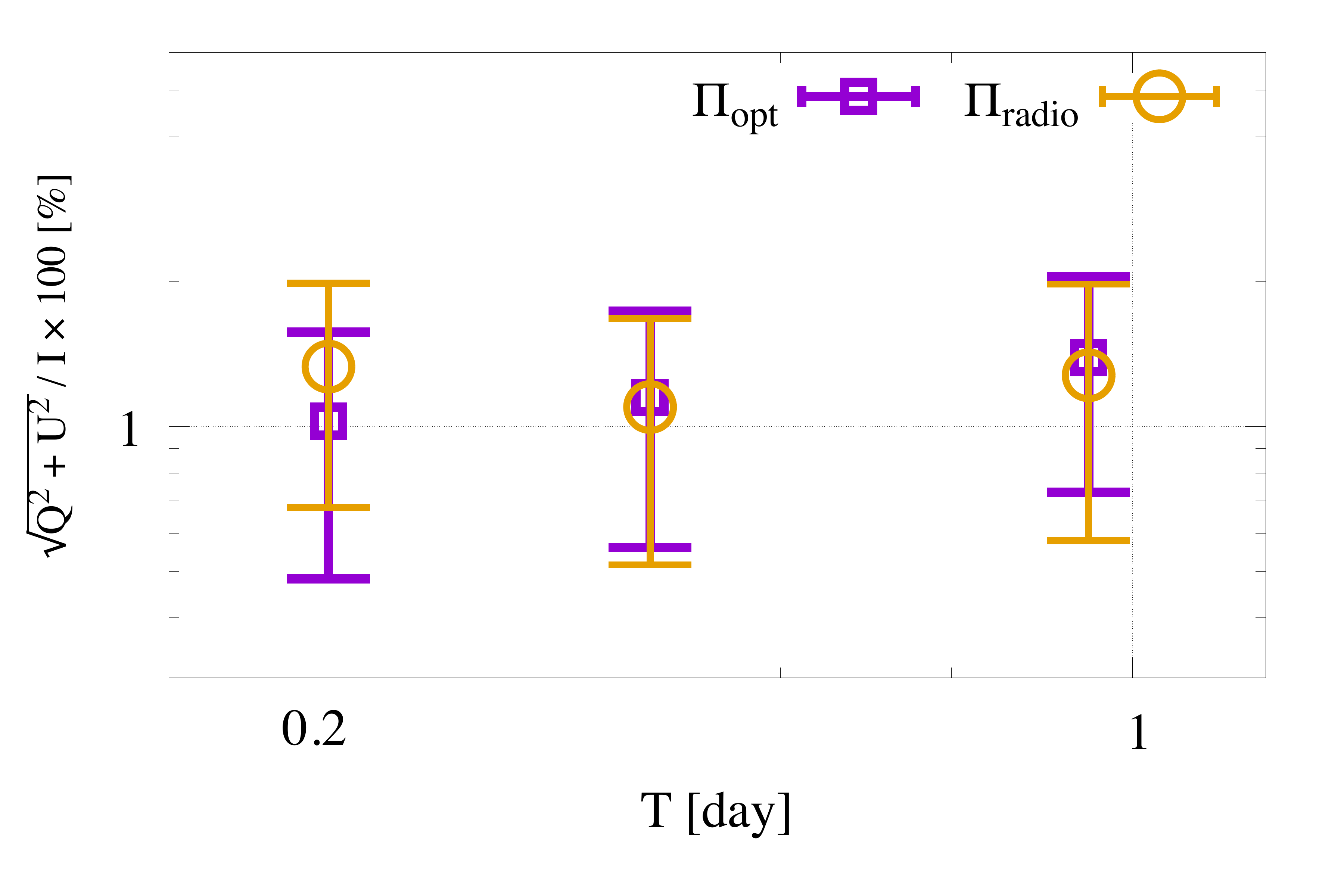}
    \caption{Realization-averaged PDs for 100 turbulent field realizations at $T=0.2,\,0.4$ and $0.9$ days for Kolmogorov spectrum. We set $f_B = 4.0$ for $k_{\rm min} = 2\pi/\lambda'_B$ and $k_{\rm max} = 8 k_{\rm min}$. 
    The marks are the same as \Figref{fig:PD-time}. \label{fig:PD_kspectrum}}
\end{figure}

We show the calculation results for $f_B = 4.0$ and $x=8$ in \Figref{fig:PD_kspectrum}. 
The average PD at each time is $\sim 4$ times lower than that calculated using only $k_{\rm min}$. This is still comparable to the observed PDs of late-time optical GRB afterglows. 
We see that $\langle \Pi_{\rm radio} \rangle \gtrsim \langle \Pi_{\rm opt} \rangle$ and that the PDs and PAs vary randomly and continuously in a similar manner as the calculation results with only $k_{\rm min}$.
We also perform calculations for $f_B = 1.0$ and find the same behaviors as $f_B = 4.0$. In summary, our conclusion in \Secref{sec:summary} does not change even if we consider the Kolmogorov-type magnetic power spectrum.

\bibliography{ms}{}
\bibliographystyle{aasjournal}

\end{document}